\documentstyle[bezier,12pt]{article}
\newcommand{\beq}{\vskip 1.5mm\begin{equation}}
\newcommand{\eeq}{\end{equation}\vskip 5.5mm\noindent}
\newcommand{\beqr}{\vskip 1.5mm\begin{eqnarray}}
\newcommand{\eeqr}{\end{eqnarray}\vskip 5.5mm\noindent}
\newcommand{\noin}{\noindent}

\renewcommand\thebibliography{\list
 {[\arabic{enumi}]}{\settowidth\labelwidth{[99]}\leftmargin\labelwidth
 \advance\leftmargin\labelsep
 \usecounter{enumi}}}

\newcommand{\ds}{\displaystyle}
\def\‹{¨­ --‹¨¢¨­£áâ®­  }
\newcommand{\rot}{\mathop{\rm rot}\nolimits}
\newcommand{\dv}{\mathop{\rm div}\nolimits}
\newcommand{\gtrsim}{\mathrel{\stackrel{\sim}{>}}}
\newcommand{\ltsim}{\mathrel{\stackrel{\sim}{<}}}
\begin{document}
%
%
%%%%%%%%%%%%%%%%%%%%%%%%%%%%%%%%%%%%%%%%%%%%%%%%%%%%%%%%%%%%%%%%%%%%%%
%                         TITLE                                      %
%%%%%%%%%%%%%%%%%%%%%%%%%%%%%%%%%%%%%%%%%%%%%%%%%%%%%%%%%%%%%%%%%%%%%%
\begin{center}

\noin
{\Large\bf Vortex loops entry into type--II
           superconductors}

\vspace{5.5mm}

%\large
\noin
A.~V.~Samokhvalov

\vspace{2.5mm}
{\small \sl
        Institute for physics of microstructures \\
	\vspace{-2mm} Russian Academy of Sciences \\
        \vspace{-2mm} 603600,\ Nizhny Novgorod, Russia \\
        \vspace{-2mm} e-mail:\ samokh@ipm.sci-nnov.ru }

\end{center}

\normalsize
%\large
\rm
\begin{abstract}
The magnetic field distribution, the magnetic flux, and the free
energy of an Abrikosov vortex loop near a flat surface of type--II
superconductors are calculated in the London approximation.
The shape of such a vortex line is a semicircle of
arbitrary radius.
The interaction of the vortex half--ring and an external homogeneous
magnetic field applied along the surface is studied.
The magnitude of the energy barrier against the vortex expansion
into superconductor is found.
The possibilities of formation of an equilibrium vortex line
determined by the structure of the applied magnetic field by
creating the expanding vortex loops near the surface
of type--II superconductor are discussed.
\end{abstract}
%
%
%%%%%%%%%%%%%%%%%%%%%%%%%%%%%%%%%%%%%%%%%%%%%%%%%%%%%%%%%%%%%%%%%%%%%%
%                        1. INTRODUCTION                             %
%%%%%%%%%%%%%%%%%%%%%%%%%%%%%%%%%%%%%%%%%%%%%%%%%%%%%%%%%%%%%%%%%%%%%%
%
\vspace{10mm}
\begin{center}
{\small\bf 1. INTRODUCTION}
\end{center}

In a mixed state a magnetic field penetrates into
type--II superconductors (SC) in the form of
Abrikosov vortices
having one quantum of magnetic flux $\Phi_0  = \pi \hbar c/e$
\cite{Abrikosov57}.
The presence of these vortices in SC and their
interaction with inhomogeneities and defects of the material
(pinning) determine both the magnetic properties of SC
\cite{Huebener&Flux}
and the ability of SC to carry a current without a resistance
\cite{Campbell&Crit}.
In the vortex core, where the magnetic field achieves its maximum,
superconductivity is destroyed and the order parameter describing
the superconducting properties of the material vanishes.
The center of the vortex core defines the location of the
vortex line (VL).
The equilibrium form of the vortex lines in type--II superconductors
reproduce the structure of an applied magnetic field.
Thus, the homogeneous magnetic field generates a two--dimensional
lattice of rectilinear vortex lines directed along the field.
The vortex penetration in SC is delayed significantly due to the
presence of the surface Bean--Livingston (BL) barrier
\cite{Bean64,Clem73}.
An increase of an applied magnetic field $H_0$
results in a decrease of the BL barrier.
The barrier is destroyed, then the field $H_0$ reaches
the thermodynamic critical field $H_{cm}$.
If the applied magnetic field $H_0 < H_{cm}$, the penetration of
vortices in SC occurs as a result of sufficiently large
fluctuations.
In this case the vortex line produced near the surface
forms a semicircle (vortex loop) the ends of which touch
the superconductor surface
\cite{Galaiko66,Petukhov73,Koshelev92}.
A minimal critical size of the loop exists so that the vortex
does not collapse at the surface but expands into the SC.

In this paper we consider the conditions of
formation of the vortex loops near a flat surface
of a type--II superconductor in the applied magnetic field
parallel to the surface.
In section 2, the magnetic field, the magnetic flux and the free
energy of the solitary Abrikosov vortex in the form of a
semicircle of arbitrary radius are calculated in the
London approach.
In section 3, the interaction of the vortex half--ring and an
applied homogeneous magnetic field is studied.
In this section we calculate the Gibbs free energy and
determine the energy barrier preventing
creation of expanding vortex loops.
In section 4 the processes of formation of equilibrium VLs,
defined by the applied magnetic field are discussed.
The development of a rectilinear VL parallel to the applied
magnetic field by intersection and cross joining of adjacent
vortex loops is considered.
%
%
%
%%%%%%%%%%%%%%%%%%%%%%%%%%%%%%%%%%%%%%%%%%%%%%%%%%%%%%%%%%%%%%%%%%%%%%
%         2. VORTEX HALF--RING NEAR THE SURFACE...                   %
%%%%%%%%%%%%%%%%%%%%%%%%%%%%%%%%%%%%%%%%%%%%%%%%%%%%%%%%%%%%%%%%%%%%%%
\vspace{10mm}
\begin{center}
{\small\bf 2. VORTEX HALF--RING NEAR THE SURFACE OF
              A SUPERCONDUCTOR}
\end{center}

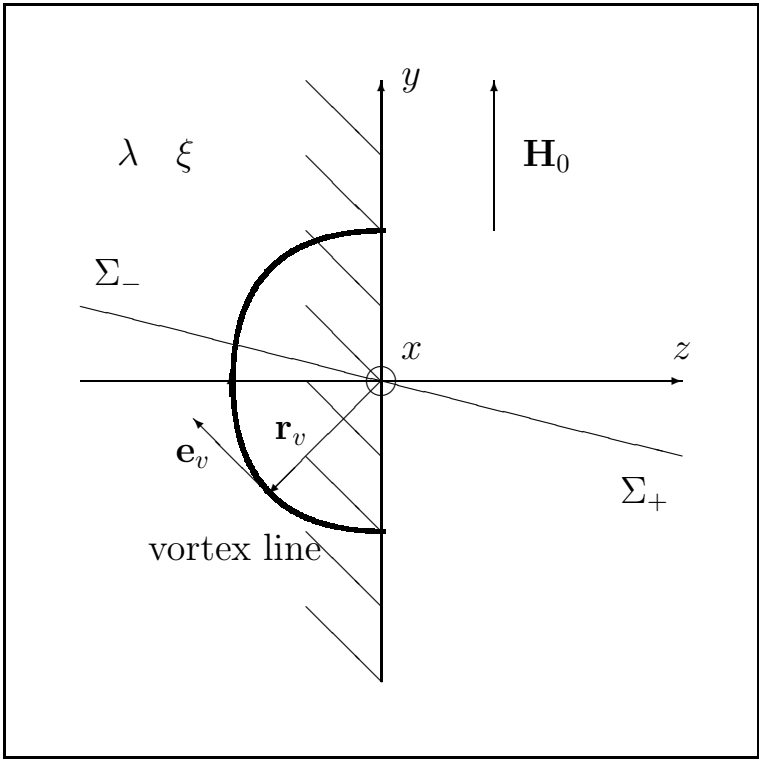
\begin{figure}
\unitlength=1.00mm
\special{em:linewidth 0.4pt}
\linethickness{0.4pt}
\begin{picture}(100.00,100)
\linethickness{0.25pt}
\linethickness{0.5pt}
\put(0.00,50.00){\framebox(100.00,100.00)[cc]{}}
\put(50.00,60.00){\vector(0,1){80.00}}
\put(10.00,100.00){\vector(1,0){80.00}}
\linethickness{1.5pt}
\bezier{160}(50.00,120.00)(30.00,120.00)(30.00,100.00)
\bezier{160}(30.00,100.00)(30.00,80.00)(50.00,80.00)
\put(30.00,98.00){\vector(0,1){3.00}}
\linethickness{0.4pt}
\put(38.00,93.00){\makebox(0,0)[cc]{${\bf r}_v$}}
\put(20.00,130.00){\makebox(0,0)[cc]{$\lambda \quad \xi$}}
\put(65.00,120.00){\vector(0,1){20.00}}
\put(72.00,130.00){\makebox(0,0)[cc]{${\bf H}_0$}}
\put(15.00,114.00){\makebox(0,0)[cc]{$\Sigma_{-}$}}
\put(85.00,85.00){\makebox(0,0)[cc]{$\Sigma_{+}$}}
\put(19.00,78.00){\makebox(0,0)[lc]{vortex line}}
\put(54.00,140.00){\makebox(0,0)[cc]{$y$}}
\put(90.00,104.00){\makebox(0,0)[cc]{$z$}}
\put(50.00,100.00){\vector(-1,-1){15.00}}
\put(40.00,140.00){\line(1,-1){10.07}}
\put(40.00,130.00){\line(1,-1){10.07}}
\put(40.00,120.00){\line(1,-1){10.07}}
\put(40.00,110.00){\line(1,-1){10.07}}
\put(40.00,100.00){\line(1,-1){10.07}}
\put(40.00,90.00){\line(1,-1){10.07}}
\put(40.00,80.00){\line(1,-1){10.07}}
\put(40.00,70.00){\line(1,-1){10.07}}
\put(10.00,110.00){\line(4,-1){80.00}}
\put(35.00,85.00){\vector(-1,1){10.00}}
\put(25.00,90.00){\makebox(0,0)[cc]{${\bf e}_v$}}
\put(50.00,100.00){\circle{4.00}}
\put(54.00,104.00){\makebox(0,0)[cc]{$x$}}
\end{picture}
\caption{Abrikosov vortex in the form of the semicircle
of the radius $\rho_v$ near the flat surface $S$ of type--II
superconductor.
The superconductor occupies the half--space $z \leq 0$ and is
characterized by the magnetic field penetration depth $\lambda$
and coherence length $\xi$; the vortex line is placed in the
plane $(yz)$.
The intersection of vortex plane and cross-section
$\Sigma = \Sigma_{-} + \Sigma_{+}$ is shown. }
\label{halfring}
\end{figure}

Let us consider a type--II superconductor occupying
the region $z \le 0$ and bounded by the surface $S$ coincident
with the coordinate plane $(xy)$.
We assume that the vortex line is a semicircle of arbitrary
radius $\rho_v$ lying in the plane $(yz)$ parallel to
the applied magnetic field ${\bf H}_0$ (fig.~\ref{halfring}).
Due to the continuity of the magnetic flux, the end points of the VL
have to be placed at the surface $S$
\cite{Galaiko66}.
The London approach is appropriate for extreme type--II
superconductors with a large Ginzburg--Landau parameter
$\kappa = \lambda / \xi \gg 1$,
where $\lambda$ is the magnetic field penetration depth
and $\xi$ is the coherence length.
In the London model the magnetic field distribution
${\bf H}$ in SC ($z \le 0$) is described by the London
equation
\cite{Abrikosov&Fund}:
\beq
{\bf H} + \rot\rot {\bf H} %
 = {\bf e}_{\it v} \delta({\bf r}-{\bf r}_{\it v}).     \label{eqnLondon} %
\eeq
Here ${\bf r}_v$ is the radius vector defining the VL location in
space, and ${\bf e}_v$ is the unit vector directed along the VL.
Outside the superconductor ($z>0$) there are no currents,
and the magnetic field ${\bf H}_s$, produced by the flux line,
is yielded by the Maxwell equations:
\beq
\rot {\bf H}_s = 0, \qquad \dv {\bf H}_s = 0 .          \label{eqnMaxwell} %
\eeq
The boundary conditions for the surface $S$ can be obtained from
the continuity of all components of the magnetic field in the plane
$z=0$:
\beq
{\bf H} = {\bf H}_s.                                    \label{BoundCond} %
\eeq
Everywhere here we use dimensionless units and assume that the
magnetic flux $\Phi$ and the coordinate vector ${\bf r}$ are scaled
by the flux quantum $\Phi_0$ and the magnetic field penetration depth
$\lambda$, respectively.
For these units the scale of the magnetic field is $\Phi_0 / \lambda^2$,
and the current density $j$ is measured by the units of
$c \Phi_0 / \lambda^3$.

Due to the linearity of equation (\ref{eqnLondon}) and
boundary conditions (\ref{BoundCond}) we represent the magnetic
field ${\bf H}$ in the SC as a superposition:
\beq
{\bf H} = {\bf H}_v + {\bf H}_d.                        \label{Superpos} %
\eeq
The term ${\bf H}_v = H_v {\bf e}_v$ describes the structure of the
magnetic field of a toroidal Abrikosov vortex with a VL consisting of
a circle of radius $\rho_v$.
In the cylindrical coordinate system
$(\rho = \sqrt{y^2+z^2}, \varphi, x)$ the component $H_v(\rho,x)$
satisfies the equation
\cite{Kozlov91}:
\beq
{\partial^2 H_v \over {\partial \rho}^2}
+ {1 \over \rho }{\partial H_v \over \partial \rho}
+ { \partial^2 H_v \over {\partial x}^2}
 - (1+ {1 \over \rho^2}) H_v
 = - \delta(\rho-\rho_v) \delta(x).                     \label{eqnLondonR} %
\eeq
The solution of equation (\ref{eqnLondonR})
for arbitrary vortex radius $\rho_v$,
was obtained in
\cite{Kozlov91,Kozlov93,Genenko94}
by means of the Fourier--Bessel transform,
and can be represented as follows:
\beq
H_v(\rho,x) = \frac{\rho_v}{2 \pi}
\int \limits_{-\infty}^{+\infty} dq \exp{(i q x)}
\cases {\ds I_1(\rho \sqrt{1+q^2}) K_1(\rho_v \sqrt{1+q^2}),
       \quad \rho \leq \rho_v ; \cr
       I_1(\rho_v \sqrt{1+q^2}) K_1(\rho \sqrt{1+q^2}),
       \quad \rho \geq \rho_v, \cr},                    \label{solLondonR} %
\eeq
where $I_1(\zeta)$, $K_1(\zeta)$ are the modified Bessel functions
of the first and second kind.
The term ${\bf H}_d$ is the solution of the homogeneous London
equation
\beq
{\bf H}_d + \rot\rot {\bf H}_{\it d} = 0               \label{eqnLondon0} %
\eeq
in the half--space $z \le 0$
with the boundary condition (\ref{BoundCond}).
To solve equation (\ref{eqnLondon0}) it is convenient to represent
the components ${\bf H}_d=(H_d^x,H_d^y,H_d^z)$ in the form
of the two dimensional Fourier transform over spatial
harmonics:
\beqr
H_d^\alpha = \frac{1}{2 \pi}
\int \limits_{-\infty}^{+\infty} dq \exp{(i q x)}
\int \limits_{-\infty}^{+\infty} du \exp{(i u y)}\,
C_d^\alpha(q,u) \exp{(z \sqrt{1+q^2+u^2})},          \label{solLondon0} \\ %
\quad \alpha = x,y,z . \nonumber %
\eeqr
Equation (\ref{eqnLondon0}) then reduces to the simple
algebraic relation for the coefficients $C_d^\alpha(q,u)$:
\beq
-q\, C_d^x - u\, C_d^y + %
   i\, \sqrt{1+q^2+u^2}\, C_d^z = 0.                    \label{relLondon0} %
\eeq
In compliance with (\ref{eqnMaxwell}), the magnetic field ${\bf H}_s$
can be written as a negative gradient of a scalar potential $U_s$:
\beq
{\bf H}_s = - \nabla U_s,                               \label{defUs} %
\eeq
and the potential $U_s$ obeys Laplace's equation
\beq
\triangle U_s = 0.                                      \label{eqnLaplace} %
\eeq
It is evident, that the solution of equation (\ref{eqnLaplace})
can be represented in the form:
\beq
U_s = \frac{1}{2 \pi}
\int \limits_{-\infty}^{+\infty} dq \exp{(i q x)}
\int \limits_{-\infty}^{+\infty} du \exp{(i u y)}
C_s(q,u) \exp{(-z \sqrt{q^2+u^2})}.                     \label{solLaplace} %
\eeq
The boundary conditions for the magnetic field (\ref{BoundCond})
lead to three algebraic equations for the so far
undetermined coefficients $C_d^\alpha(q,u)$, $C_s(q,u)$:
\beq
C_d^x = -i\, q\, C_s, \quad
C_d^y = -i\, u\, C_s, \quad
C_d^z - \frac{i}{\pi} S_v(q,u)= \sqrt{q^2+u^2}\, C_s.   \label{relBoundCond} %
\eeq
The solutions of the linear algebraic system
(\ref{relLondon0},\ref{relBoundCond})
for $q^2+u^2 \neq 0$ are
\beqr
& &C_d^x = -\frac{\ds q \sqrt{1+q^2+u^2}\: S_v(q,u)}
        {\ds \pi \sqrt{q^2+u^2}\:
        \left( \sqrt{1+q^2+u^2}+\sqrt{q^2+u^2} \right) },\nonumber \\
& &C_d^y = -\frac{\ds u \sqrt{1+q^2+u^2}\: S_v(q,u)}
        {\ds \pi \sqrt{q^2+u^2}\:
        \left( \sqrt{1+q^2+u^2}+\sqrt{q^2+u^2} \right) },\label{solC} \\ %
& &C_d^z = \frac{\ds i \sqrt{q^2+u^2}\: S_v(q,u)}
        {\ds \pi \left( \sqrt{1+q^2+u^2}
                        +\sqrt{q^2+u^2} \right) }, \nonumber \\
& &C_s = -\frac{\ds i \sqrt{1+q^2+u^2}\: S_v(q,u)}
        {\ds \pi \sqrt{q^2+u^2}\:
        \left( \sqrt{1+q^2+u^2}+\sqrt{q^2+u^2} \right) }.\nonumber
\eeqr
If $q^2+u^2 = 0$, then the coefficients $C_d^\alpha$, $C_s$ in the
expansions (\ref{solLondon0},\ref{solLaplace}) are equal to zero.
The function $S_v(q,u)$ is determined by the spatial Fourier
spectrum of the magnetic field distribution $H_v$
in the plane $z=0$:
\beq
S_v(q,u) ={i \over 2 } %
\int\limits_{-\infty}^{+\infty} dx \exp{(-i q x)} %
\int\limits_{-\infty}^{+\infty} dy
        \exp{(-i u y)} H_v(y,x),                        \label{FourSpec} %
\eeq
Substitution of $H_v$ (\ref{solLondonR}) in the expansion
(\ref{FourSpec}) leads to the following representation of the
spectral function  $S_v(q,u)$:
\beq
S_v(q,u) = \rho_v \int\limits_0^{+\infty} dy \sin{(u y)}
\cases {\ds I_1(y \sqrt{1+q^2}) K_1(\rho_v \sqrt{1+q^2}),
       \quad y \leq \rho_v ; \cr
       I_1(\rho_v \sqrt{1+q^2}) K_1(y \sqrt{1+q^2}),
       \quad y \geq \rho_v . \cr}                       \label{SpecFunc} %
\eeq
The details of the numerical simulations of the spectral function
$S_v(q,u)$ are presented in Appendix~A.
Using the asymptotic expansion of the modified Bessel functions
$I_1(\zeta)$, $K_1(\zeta)$ for $\zeta \gg 1$ one can obtain
\beq
S_v(q,u) \sim \frac{\ds \sin{(u \rho_v)}}{\ds u^2},
       \quad u \gg 1, \quad u \gg q,                    \label{SpecFuncAs} %
\eeq
which is a good approximation of the exact expression
(\ref{SpecFunc}) at the large values of spectral variable $u$.
In the other limiting case ( $q,u \ll 1$, $u\rho_v \ll 1$ )
it is convenient to represent the spectral function
$S_v(q,u)$ (\ref{SpecFunc})  as
\beq
S_v(q,u) \simeq a\, u ,
       \qquad q,u \ll 1, \quad u\rho_v \ll 1,           \label{SpecFuncAs1} %
\eeq
where the factor $a$ is expressed in terms of
the vortex radius $\rho_v$ :
\beq
a = \rho_v %
\left[ %
  K_1(\rho_v) \int\limits_{0} ^{\rho_v} d\zeta\, \zeta I_1(\zeta) %
 +I_1(\rho_v) \int\limits_{\rho_v} ^{+\infty} d\zeta\, \zeta K_1(\zeta) %
\right].                                                \label{FactorA} %
\eeq

Thus, the magnetic field structure of a solitary vortex half--ring
placed near the flat surface of a type--II superconductor is fully
determined by the relations
(\ref{Superpos},\ref{solLondonR},\ref{solLondon0},\ref{defUs},%
\ref{solLaplace},\ref{solC},\ref{SpecFunc}).
The field ${\bf H}_d$ describes the disturbances of the structure
of the toroidal Abrikosov vortex (\ref{solLondonR}) by the
surface $S$.
In the area of the vortex core the field ${\bf H}_d$ is in opposite
direction to the vortex line, so that
$\left( {\bf H}_d \cdot {\bf H}_v \right) < 0$.
It results in a decrease of the magnetic field value at the VL.
The superconductor surface effect is strong only in
the surface layer of a thickness defined
by the vortex radius $\rho_v$ or the magnetic
field penetration depth $\lambda$.

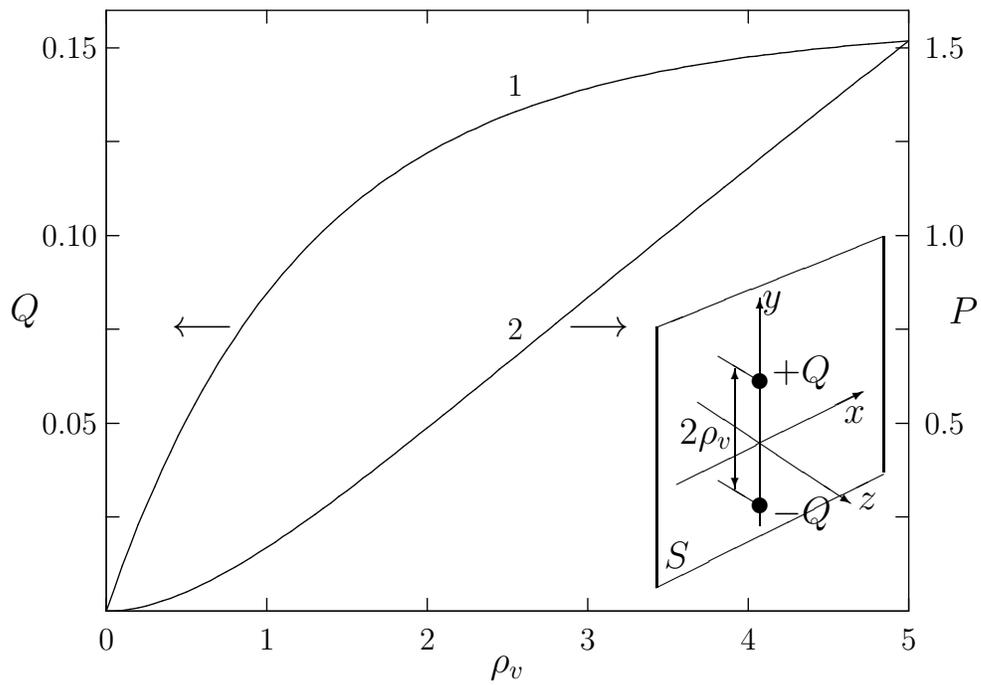
\begin{figure}[htb]
%\input{dipole.pic}
% GNUPLOT: LaTeX picture with emtex specials
\setlength{\unitlength}{0.240900pt}
\ifx\plotpoint\undefined\newsavebox{\plotpoint}\fi
\sbox{\plotpoint}{\rule[-0.500pt]{1.000pt}{1.000pt}}%
\special{em:linewidth 0.5pt}%
\begin{picture}(1500,1080)(0,0)
\font\gnuplot=cmr12 at 12pt
\gnuplot
\put(176,113){\special{em:moveto}}
\put(1436,113){\special{em:lineto}}
\put(176,113){\special{em:moveto}}
\put(176,1057){\special{em:lineto}}
\put(176,261){\special{em:moveto}}
\put(196,261){\special{em:lineto}}
\put(1436,261){\special{em:moveto}}
\put(1416,261){\special{em:lineto}}
\put(154,261){\makebox(0,0)[r]{ }}
\put(176,408){\special{em:moveto}}
\put(196,408){\special{em:lineto}}
\put(1436,408){\special{em:moveto}}
\put(1416,408){\special{em:lineto}}
\put(154,408){\makebox(0,0)[r]{ }}
\put(176,556){\special{em:moveto}}
\put(196,556){\special{em:lineto}}
\put(1436,556){\special{em:moveto}}
\put(1416,556){\special{em:lineto}}
\put(154,556){\makebox(0,0)[r]{ }}
\put(176,703){\special{em:moveto}}
\put(196,703){\special{em:lineto}}
\put(1436,703){\special{em:moveto}}
\put(1416,703){\special{em:lineto}}
\put(154,703){\makebox(0,0)[r]{ }}
\put(176,851){\special{em:moveto}}
\put(196,851){\special{em:lineto}}
\put(1436,851){\special{em:moveto}}
\put(1416,851){\special{em:lineto}}
\put(154,851){\makebox(0,0)[r]{ }}
\put(176,998){\special{em:moveto}}
\put(196,998){\special{em:lineto}}
\put(1436,998){\special{em:moveto}}
\put(1416,998){\special{em:lineto}}
\put(154,998){\makebox(0,0)[r]{ }}
\put(176,113){\special{em:moveto}}
\put(176,133){\special{em:lineto}}
\put(176,1057){\special{em:moveto}}
\put(176,1037){\special{em:lineto}}
\put(176,68){\makebox(0,0){0}}
\put(428,113){\special{em:moveto}}
\put(428,133){\special{em:lineto}}
\put(428,1057){\special{em:moveto}}
\put(428,1037){\special{em:lineto}}
\put(428,68){\makebox(0,0){1}}
\put(680,113){\special{em:moveto}}
\put(680,133){\special{em:lineto}}
\put(680,1057){\special{em:moveto}}
\put(680,1037){\special{em:lineto}}
\put(680,68){\makebox(0,0){2}}
\put(932,113){\special{em:moveto}}
\put(932,133){\special{em:lineto}}
\put(932,1057){\special{em:moveto}}
\put(932,1037){\special{em:lineto}}
\put(932,68){\makebox(0,0){3}}
\put(1184,113){\special{em:moveto}}
\put(1184,133){\special{em:lineto}}
\put(1184,1057){\special{em:moveto}}
\put(1184,1037){\special{em:lineto}}
\put(1184,68){\makebox(0,0){4}}
\put(1436,113){\special{em:moveto}}
\put(1436,133){\special{em:lineto}}
\put(1436,1057){\special{em:moveto}}
\put(1436,1037){\special{em:lineto}}
\put(1436,68){\makebox(0,0){5}}
\put(176,113){\special{em:moveto}}
\put(1436,113){\special{em:lineto}}
\put(1436,1057){\special{em:lineto}}
\put(176,1057){\special{em:lineto}}
\put(176,113){\special{em:lineto}}
\put(806,23){\makebox(0,0){$\rho_v$}}
\put(25,585){\makebox(0,0)[l]{$Q$}}
\put(806,556){\makebox(0,0)[l]{2}}
\put(806,939){\makebox(0,0)[l]{1}}
\put(1499,585){\makebox(0,0)[l]{$P$}}
\put(75,408){\makebox(0,0)[l]{0.05}}
\put(75,703){\makebox(0,0)[l]{0.10}}
\put(75,998){\makebox(0,0)[l]{0.15}}
\put(1461,408){\makebox(0,0)[l]{0.5}}
\put(1461,703){\makebox(0,0)[l]{1.0}}
\put(1461,998){\makebox(0,0)[l]{1.5}}
\put(900,556){\makebox(0,0)[l]{$\longrightarrow$}}
\put(375,556){\makebox(0,0)[r]{$\longleftarrow$}}
\put(176,113){\special{em:moveto}}
\put(189,149){\special{em:lineto}}
\put(201,184){\special{em:lineto}}
\put(214,217){\special{em:lineto}}
\put(226,249){\special{em:lineto}}
\put(239,279){\special{em:lineto}}
\put(252,308){\special{em:lineto}}
\put(264,336){\special{em:lineto}}
\put(277,363){\special{em:lineto}}
\put(289,389){\special{em:lineto}}
\put(302,414){\special{em:lineto}}
\put(315,438){\special{em:lineto}}
\put(327,460){\special{em:lineto}}
\put(340,482){\special{em:lineto}}
\put(352,503){\special{em:lineto}}
\put(365,523){\special{em:lineto}}
\put(378,542){\special{em:lineto}}
\put(390,561){\special{em:lineto}}
\put(403,579){\special{em:lineto}}
\put(415,596){\special{em:lineto}}
\put(428,612){\special{em:lineto}}
\put(441,628){\special{em:lineto}}
\put(453,643){\special{em:lineto}}
\put(466,657){\special{em:lineto}}
\put(478,671){\special{em:lineto}}
\put(491,685){\special{em:lineto}}
\put(504,698){\special{em:lineto}}
\put(516,710){\special{em:lineto}}
\put(529,722){\special{em:lineto}}
\put(541,733){\special{em:lineto}}
\put(554,744){\special{em:lineto}}
\put(567,755){\special{em:lineto}}
\put(579,765){\special{em:lineto}}
\put(592,774){\special{em:lineto}}
\put(604,784){\special{em:lineto}}
\put(617,793){\special{em:lineto}}
\put(630,801){\special{em:lineto}}
\put(642,810){\special{em:lineto}}
\put(655,818){\special{em:lineto}}
\put(667,825){\special{em:lineto}}
\put(680,833){\special{em:lineto}}
\put(693,840){\special{em:lineto}}
\put(705,847){\special{em:lineto}}
\put(718,853){\special{em:lineto}}
\put(730,859){\special{em:lineto}}
\put(743,866){\special{em:lineto}}
\put(756,871){\special{em:lineto}}
\put(768,877){\special{em:lineto}}
\put(781,883){\special{em:lineto}}
\put(793,888){\special{em:lineto}}
\put(806,893){\special{em:lineto}}
\put(819,898){\special{em:lineto}}
\put(831,902){\special{em:lineto}}
\put(844,907){\special{em:lineto}}
\put(856,911){\special{em:lineto}}
\put(869,915){\special{em:lineto}}
\put(882,919){\special{em:lineto}}
\put(894,923){\special{em:lineto}}
\put(907,927){\special{em:lineto}}
\put(919,931){\special{em:lineto}}
\put(932,934){\special{em:lineto}}
\put(945,938){\special{em:lineto}}
\put(957,941){\special{em:lineto}}
\put(970,944){\special{em:lineto}}
\put(982,947){\special{em:lineto}}
\put(995,950){\special{em:lineto}}
\put(1008,953){\special{em:lineto}}
\put(1020,955){\special{em:lineto}}
\put(1033,958){\special{em:lineto}}
\put(1045,961){\special{em:lineto}}
\put(1058,963){\special{em:lineto}}
\put(1071,965){\special{em:lineto}}
\put(1083,968){\special{em:lineto}}
\put(1096,970){\special{em:lineto}}
\put(1108,972){\special{em:lineto}}
\put(1121,974){\special{em:lineto}}
\put(1134,976){\special{em:lineto}}
\put(1146,978){\special{em:lineto}}
\put(1159,980){\special{em:lineto}}
\put(1171,982){\special{em:lineto}}
\put(1184,984){\special{em:lineto}}
\put(1197,985){\special{em:lineto}}
\put(1209,987){\special{em:lineto}}
\put(1222,988){\special{em:lineto}}
\put(1234,990){\special{em:lineto}}
\put(1247,991){\special{em:lineto}}
\put(1260,993){\special{em:lineto}}
\put(1272,994){\special{em:lineto}}
\put(1285,996){\special{em:lineto}}
\put(1297,997){\special{em:lineto}}
\put(1310,998){\special{em:lineto}}
\put(1323,999){\special{em:lineto}}
\put(1335,1001){\special{em:lineto}}
\put(1348,1002){\special{em:lineto}}
\put(1360,1003){\special{em:lineto}}
\put(1373,1004){\special{em:lineto}}
\put(1386,1005){\special{em:lineto}}
\put(1398,1006){\special{em:lineto}}
\put(1411,1007){\special{em:lineto}}
\put(1423,1008){\special{em:lineto}}
\put(1436,1009){\special{em:lineto}}
\put(176,113){\special{em:moveto}}
\put(189,113){\special{em:lineto}}
\put(201,114){\special{em:lineto}}
\put(214,116){\special{em:lineto}}
\put(226,118){\special{em:lineto}}
\put(239,121){\special{em:lineto}}
\put(252,125){\special{em:lineto}}
\put(264,129){\special{em:lineto}}
\put(277,133){\special{em:lineto}}
\put(289,138){\special{em:lineto}}
\put(302,143){\special{em:lineto}}
\put(315,149){\special{em:lineto}}
\put(327,155){\special{em:lineto}}
\put(340,161){\special{em:lineto}}
\put(352,168){\special{em:lineto}}
\put(365,175){\special{em:lineto}}
\put(378,182){\special{em:lineto}}
\put(390,189){\special{em:lineto}}
\put(403,197){\special{em:lineto}}
\put(415,205){\special{em:lineto}}
\put(428,213){\special{em:lineto}}
\put(441,221){\special{em:lineto}}
\put(453,230){\special{em:lineto}}
\put(466,238){\special{em:lineto}}
\put(478,247){\special{em:lineto}}
\put(491,256){\special{em:lineto}}
\put(504,265){\special{em:lineto}}
\put(516,274){\special{em:lineto}}
\put(529,283){\special{em:lineto}}
\put(541,293){\special{em:lineto}}
\put(554,302){\special{em:lineto}}
\put(567,312){\special{em:lineto}}
\put(579,322){\special{em:lineto}}
\put(592,331){\special{em:lineto}}
\put(604,341){\special{em:lineto}}
\put(617,351){\special{em:lineto}}
\put(630,361){\special{em:lineto}}
\put(642,371){\special{em:lineto}}
\put(655,381){\special{em:lineto}}
\put(667,391){\special{em:lineto}}
\put(680,401){\special{em:lineto}}
\put(693,411){\special{em:lineto}}
\put(705,421){\special{em:lineto}}
\put(718,431){\special{em:lineto}}
\put(730,441){\special{em:lineto}}
\put(743,452){\special{em:lineto}}
\put(756,462){\special{em:lineto}}
\put(768,472){\special{em:lineto}}
\put(781,482){\special{em:lineto}}
\put(793,493){\special{em:lineto}}
\put(806,503){\special{em:lineto}}
\put(819,513){\special{em:lineto}}
\put(831,523){\special{em:lineto}}
\put(844,534){\special{em:lineto}}
\put(856,544){\special{em:lineto}}
\put(869,554){\special{em:lineto}}
\put(882,565){\special{em:lineto}}
\put(894,575){\special{em:lineto}}
\put(907,585){\special{em:lineto}}
\put(919,595){\special{em:lineto}}
\put(932,606){\special{em:lineto}}
\put(945,616){\special{em:lineto}}
\put(957,626){\special{em:lineto}}
\put(970,637){\special{em:lineto}}
\put(982,647){\special{em:lineto}}
\put(995,657){\special{em:lineto}}
\put(1008,667){\special{em:lineto}}
\put(1020,677){\special{em:lineto}}
\put(1033,688){\special{em:lineto}}
\put(1045,698){\special{em:lineto}}
\put(1058,708){\special{em:lineto}}
\put(1071,718){\special{em:lineto}}
\put(1083,728){\special{em:lineto}}
\put(1096,739){\special{em:lineto}}
\put(1108,749){\special{em:lineto}}
\put(1121,759){\special{em:lineto}}
\put(1134,769){\special{em:lineto}}
\put(1146,779){\special{em:lineto}}
\put(1159,789){\special{em:lineto}}
\put(1171,799){\special{em:lineto}}
\put(1184,809){\special{em:lineto}}
\put(1197,820){\special{em:lineto}}
\put(1209,830){\special{em:lineto}}
\put(1222,840){\special{em:lineto}}
\put(1234,850){\special{em:lineto}}
\put(1247,860){\special{em:lineto}}
\put(1260,870){\special{em:lineto}}
\put(1272,880){\special{em:lineto}}
\put(1285,890){\special{em:lineto}}
\put(1297,900){\special{em:lineto}}
\put(1310,910){\special{em:lineto}}
\put(1323,920){\special{em:lineto}}
\put(1335,930){\special{em:lineto}}
\put(1348,940){\special{em:lineto}}
\put(1360,950){\special{em:lineto}}
\put(1373,959){\special{em:lineto}}
\put(1386,969){\special{em:lineto}}
\put(1398,979){\special{em:lineto}}
\put(1411,989){\special{em:lineto}}
\put(1423,999){\special{em:lineto}}
\put(1436,1009){\special{em:lineto}}
\end{picture}
\caption{Effective charge $Q$ (1) %
and dipole moment $P=2 \rho_v Q$ (2) %
vs. radius of the vortex line $\rho_v$. %
The inset shows the dipole $P$ placed in the %
plane $z=0$. } %
\label{dipole}
\end{figure}

The field ${\bf H}_s$ denotes the scattered
magnetic field outside the superconductor, produced by the
isolated flux line in the form of the half--ring placed near
the surface.
To determine a structure of the field ${\bf H}_s$ let us
estimate a behavior of the potential
$U_s$ (\ref{solLaplace}) at long distances
$z \gg 1,\, \rho_v$ from the surface.
Since the integrand in (\ref{solLaplace}) drops abruptly
if $q^2+u^2 > {(1/z)}^2$, only small values of the spectral
variables $q$ and $u$ are important.
Assuming that $q,\, u \ll 1,\, 1/ \rho_v$ and using the
approximate expression of the spectral function
$S_v(q,u)$ (\ref{SpecFuncAs1}),
the potential $U_s$ (\ref{solLaplace})
can be written in the form:
\beq
U_s \simeq \frac{2 a}{\pi^2}
\int \limits_0^{+\infty} dq \cos{(q x)}
\int \limits_0^{+\infty} du
\frac{\ds u \sin{(u y)}}{\ds \sqrt{q^2+u^2}}
\exp{(-z \sqrt{q^2+u^2})},
\qquad z \gg 1,\; \rho_v.                               \label{UsAsympt} %
\eeq
It is shown in Appendix~B, that the potential (\ref{UsAsympt})
is the potential of the dipole $P=2 \rho_v Q$
(see the inset of fig.~\ref{dipole}),
if the effective charge $Q=a/2 \pi \rho_v$.
Figure~\ref{dipole} plots the values of the charge $Q$
and dipole moment $P$ versus
the vortex radius $\rho_v$.
The dipole structure of the potential $U_s$
determines the asymptotic decay law
for the scattered field ${\bf H}_s$
at long distances $z \gg 1,\, \rho_v$ from the surface:
$H_s \sim R^{-3}$,
where $R={(x^2+y^2+z^2)}^{1/2} \gg \rho_v$
is the distance from the center of the vortex half--loop.

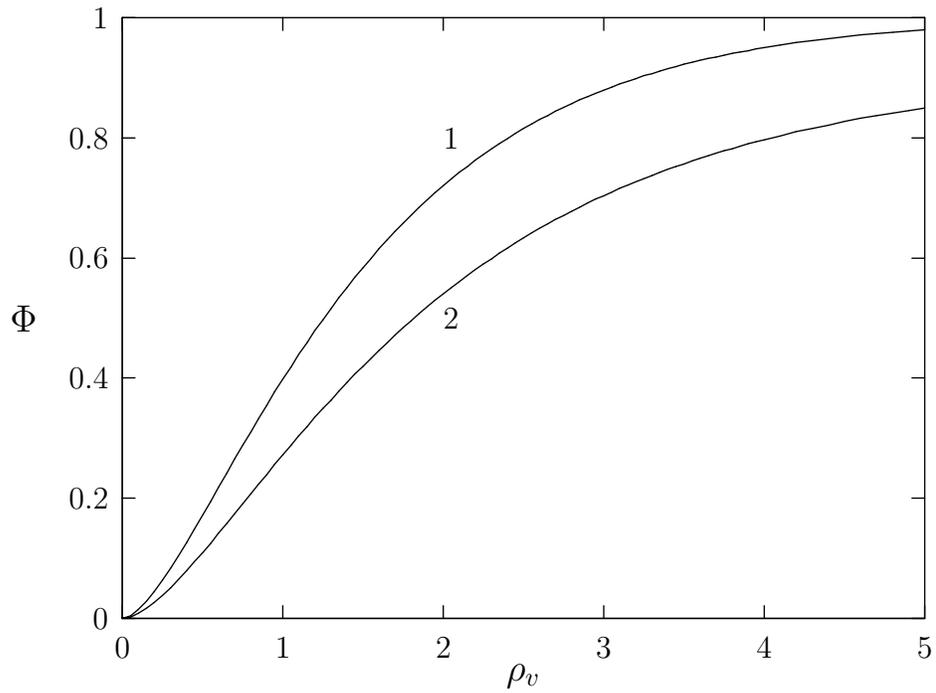
\begin{figure}[htb]
%\input{flux.pic}
% GNUPLOT: LaTeX picture with emtex specials
\setlength{\unitlength}{0.240900pt}
\ifx\plotpoint\undefined\newsavebox{\plotpoint}\fi
\sbox{\plotpoint}{\rule[-0.500pt]{1.000pt}{1.000pt}}%
\special{em:linewidth 0.5pt}%
\begin{picture}(1500,1080)(0,0)
\font\gnuplot=cmr12 at 12pt
\gnuplot
\put(176,113){\special{em:moveto}}
\put(1436,113){\special{em:lineto}}
\put(176,113){\special{em:moveto}}
\put(176,1057){\special{em:lineto}}
\put(176,113){\special{em:moveto}}
\put(196,113){\special{em:lineto}}
\put(1436,113){\special{em:moveto}}
\put(1416,113){\special{em:lineto}}
\put(154,113){\makebox(0,0)[r]{0}}
\put(176,302){\special{em:moveto}}
\put(196,302){\special{em:lineto}}
\put(1436,302){\special{em:moveto}}
\put(1416,302){\special{em:lineto}}
\put(154,302){\makebox(0,0)[r]{0.2}}
\put(176,491){\special{em:moveto}}
\put(196,491){\special{em:lineto}}
\put(1436,491){\special{em:moveto}}
\put(1416,491){\special{em:lineto}}
\put(154,491){\makebox(0,0)[r]{0.4}}
\put(176,679){\special{em:moveto}}
\put(196,679){\special{em:lineto}}
\put(1436,679){\special{em:moveto}}
\put(1416,679){\special{em:lineto}}
\put(154,679){\makebox(0,0)[r]{0.6}}
\put(176,868){\special{em:moveto}}
\put(196,868){\special{em:lineto}}
\put(1436,868){\special{em:moveto}}
\put(1416,868){\special{em:lineto}}
\put(154,868){\makebox(0,0)[r]{0.8}}
\put(176,1057){\special{em:moveto}}
\put(196,1057){\special{em:lineto}}
\put(1436,1057){\special{em:moveto}}
\put(1416,1057){\special{em:lineto}}
\put(154,1057){\makebox(0,0)[r]{1}}
\put(176,113){\special{em:moveto}}
\put(176,133){\special{em:lineto}}
\put(176,1057){\special{em:moveto}}
\put(176,1037){\special{em:lineto}}
\put(176,68){\makebox(0,0){0}}
\put(428,113){\special{em:moveto}}
\put(428,133){\special{em:lineto}}
\put(428,1057){\special{em:moveto}}
\put(428,1037){\special{em:lineto}}
\put(428,68){\makebox(0,0){1}}
\put(680,113){\special{em:moveto}}
\put(680,133){\special{em:lineto}}
\put(680,1057){\special{em:moveto}}
\put(680,1037){\special{em:lineto}}
\put(680,68){\makebox(0,0){2}}
\put(932,113){\special{em:moveto}}
\put(932,133){\special{em:lineto}}
\put(932,1057){\special{em:moveto}}
\put(932,1037){\special{em:lineto}}
\put(932,68){\makebox(0,0){3}}
\put(1184,113){\special{em:moveto}}
\put(1184,133){\special{em:lineto}}
\put(1184,1057){\special{em:moveto}}
\put(1184,1037){\special{em:lineto}}
\put(1184,68){\makebox(0,0){4}}
\put(1436,113){\special{em:moveto}}
\put(1436,133){\special{em:lineto}}
\put(1436,1057){\special{em:moveto}}
\put(1436,1037){\special{em:lineto}}
\put(1436,68){\makebox(0,0){5}}
\put(176,113){\special{em:moveto}}
\put(1436,113){\special{em:lineto}}
\put(1436,1057){\special{em:lineto}}
\put(176,1057){\special{em:lineto}}
\put(176,113){\special{em:lineto}}
\put(806,23){\makebox(0,0){$\rho_v$}}
\put(680,868){\makebox(0,0)[l]{1}}
\put(680,585){\makebox(0,0)[l]{2}}
\put(0,585){\makebox(0,0)[l]{$\Phi$}}
\put(176,113){\special{em:moveto}}
\put(189,117){\special{em:lineto}}
\put(201,127){\special{em:lineto}}
\put(214,140){\special{em:lineto}}
\put(226,155){\special{em:lineto}}
\put(239,173){\special{em:lineto}}
\put(252,192){\special{em:lineto}}
\put(264,211){\special{em:lineto}}
\put(277,232){\special{em:lineto}}
\put(289,253){\special{em:lineto}}
\put(302,275){\special{em:lineto}}
\put(315,297){\special{em:lineto}}
\put(327,319){\special{em:lineto}}
\put(340,341){\special{em:lineto}}
\put(352,363){\special{em:lineto}}
\put(365,385){\special{em:lineto}}
\put(378,406){\special{em:lineto}}
\put(390,427){\special{em:lineto}}
\put(403,448){\special{em:lineto}}
\put(415,469){\special{em:lineto}}
\put(428,489){\special{em:lineto}}
\put(441,508){\special{em:lineto}}
\put(453,528){\special{em:lineto}}
\put(466,546){\special{em:lineto}}
\put(478,565){\special{em:lineto}}
\put(491,582){\special{em:lineto}}
\put(504,600){\special{em:lineto}}
\put(516,617){\special{em:lineto}}
\put(529,633){\special{em:lineto}}
\put(541,649){\special{em:lineto}}
\put(554,664){\special{em:lineto}}
\put(567,679){\special{em:lineto}}
\put(579,694){\special{em:lineto}}
\put(592,708){\special{em:lineto}}
\put(604,721){\special{em:lineto}}
\put(617,734){\special{em:lineto}}
\put(630,747){\special{em:lineto}}
\put(642,759){\special{em:lineto}}
\put(655,771){\special{em:lineto}}
\put(667,782){\special{em:lineto}}
\put(680,793){\special{em:lineto}}
\put(693,804){\special{em:lineto}}
\put(705,814){\special{em:lineto}}
\put(718,823){\special{em:lineto}}
\put(730,833){\special{em:lineto}}
\put(743,842){\special{em:lineto}}
\put(756,851){\special{em:lineto}}
\put(768,859){\special{em:lineto}}
\put(781,867){\special{em:lineto}}
\put(793,875){\special{em:lineto}}
\put(806,883){\special{em:lineto}}
\put(819,890){\special{em:lineto}}
\put(831,897){\special{em:lineto}}
\put(844,903){\special{em:lineto}}
\put(856,910){\special{em:lineto}}
\put(869,916){\special{em:lineto}}
\put(882,922){\special{em:lineto}}
\put(894,928){\special{em:lineto}}
\put(907,933){\special{em:lineto}}
\put(919,938){\special{em:lineto}}
\put(932,943){\special{em:lineto}}
\put(945,948){\special{em:lineto}}
\put(957,953){\special{em:lineto}}
\put(970,957){\special{em:lineto}}
\put(982,961){\special{em:lineto}}
\put(995,966){\special{em:lineto}}
\put(1008,969){\special{em:lineto}}
\put(1020,973){\special{em:lineto}}
\put(1033,977){\special{em:lineto}}
\put(1045,980){\special{em:lineto}}
\put(1058,984){\special{em:lineto}}
\put(1071,987){\special{em:lineto}}
\put(1083,990){\special{em:lineto}}
\put(1096,993){\special{em:lineto}}
\put(1108,995){\special{em:lineto}}
\put(1121,998){\special{em:lineto}}
\put(1134,1001){\special{em:lineto}}
\put(1146,1003){\special{em:lineto}}
\put(1159,1005){\special{em:lineto}}
\put(1171,1008){\special{em:lineto}}
\put(1184,1010){\special{em:lineto}}
\put(1209,1014){\special{em:lineto}}
\put(1234,1018){\special{em:lineto}}
\put(1260,1021){\special{em:lineto}}
\put(1285,1024){\special{em:lineto}}
\put(1310,1027){\special{em:lineto}}
\put(1335,1030){\special{em:lineto}}
\put(1360,1032){\special{em:lineto}}
\put(1386,1034){\special{em:lineto}}
\put(1411,1036){\special{em:lineto}}
\put(1436,1038){\special{em:lineto}}
\put(176,113){\special{em:moveto}}
\put(189,115){\special{em:lineto}}
\put(201,121){\special{em:lineto}}
\put(214,129){\special{em:lineto}}
\put(226,138){\special{em:lineto}}
\put(239,149){\special{em:lineto}}
\put(252,161){\special{em:lineto}}
\put(264,174){\special{em:lineto}}
\put(277,188){\special{em:lineto}}
\put(289,202){\special{em:lineto}}
\put(302,216){\special{em:lineto}}
\put(315,231){\special{em:lineto}}
\put(327,247){\special{em:lineto}}
\put(340,262){\special{em:lineto}}
\put(352,277){\special{em:lineto}}
\put(365,293){\special{em:lineto}}
\put(378,309){\special{em:lineto}}
\put(390,324){\special{em:lineto}}
\put(403,339){\special{em:lineto}}
\put(415,355){\special{em:lineto}}
\put(428,370){\special{em:lineto}}
\put(441,385){\special{em:lineto}}
\put(453,400){\special{em:lineto}}
\put(466,414){\special{em:lineto}}
\put(478,429){\special{em:lineto}}
\put(491,443){\special{em:lineto}}
\put(504,456){\special{em:lineto}}
\put(516,470){\special{em:lineto}}
\put(529,484){\special{em:lineto}}
\put(541,497){\special{em:lineto}}
\put(554,509){\special{em:lineto}}
\put(567,522){\special{em:lineto}}
\put(579,534){\special{em:lineto}}
\put(592,546){\special{em:lineto}}
\put(604,558){\special{em:lineto}}
\put(617,570){\special{em:lineto}}
\put(630,581){\special{em:lineto}}
\put(642,592){\special{em:lineto}}
\put(655,603){\special{em:lineto}}
\put(667,613){\special{em:lineto}}
\put(680,623){\special{em:lineto}}
\put(693,633){\special{em:lineto}}
\put(705,642){\special{em:lineto}}
\put(718,652){\special{em:lineto}}
\put(730,661){\special{em:lineto}}
\put(743,670){\special{em:lineto}}
\put(756,678){\special{em:lineto}}
\put(768,687){\special{em:lineto}}
\put(781,695){\special{em:lineto}}
\put(793,703){\special{em:lineto}}
\put(806,711){\special{em:lineto}}
\put(819,719){\special{em:lineto}}
\put(831,726){\special{em:lineto}}
\put(844,733){\special{em:lineto}}
\put(856,740){\special{em:lineto}}
\put(869,746){\special{em:lineto}}
\put(882,753){\special{em:lineto}}
\put(894,759){\special{em:lineto}}
\put(907,766){\special{em:lineto}}
\put(919,772){\special{em:lineto}}
\put(932,777){\special{em:lineto}}
\put(945,783){\special{em:lineto}}
\put(957,789){\special{em:lineto}}
\put(970,794){\special{em:lineto}}
\put(982,799){\special{em:lineto}}
\put(995,804){\special{em:lineto}}
\put(1008,809){\special{em:lineto}}
\put(1020,814){\special{em:lineto}}
\put(1033,819){\special{em:lineto}}
\put(1045,823){\special{em:lineto}}
\put(1058,827){\special{em:lineto}}
\put(1071,832){\special{em:lineto}}
\put(1083,836){\special{em:lineto}}
\put(1096,840){\special{em:lineto}}
\put(1108,844){\special{em:lineto}}
\put(1121,848){\special{em:lineto}}
\put(1134,851){\special{em:lineto}}
\put(1146,855){\special{em:lineto}}
\put(1159,859){\special{em:lineto}}
\put(1171,862){\special{em:lineto}}
\put(1184,865){\special{em:lineto}}
\put(1209,871){\special{em:lineto}}
\put(1234,878){\special{em:lineto}}
\put(1260,883){\special{em:lineto}}
\put(1285,888){\special{em:lineto}}
\put(1310,894){\special{em:lineto}}
\put(1335,899){\special{em:lineto}}
\put(1360,903){\special{em:lineto}}
\put(1386,907){\special{em:lineto}}
\put(1411,911){\special{em:lineto}}
\put(1436,915){\special{em:lineto}}
\end{picture}
\caption{Magnetic flux $\Phi_v$ of toroidal Abrikosov vortex (1) %
and magnetic flux $\Phi_{\Sigma_{\ds -}}$ of the vortex %
half--ring (2) vs. radius of the vortex line $\rho_v$. } %
\label{flux}
\end{figure}

To calculate the magnetic flux $\Phi$ in an Abrikosov vortex,
the cross--section $\Sigma$ has to be selected perpendicular
to the vortex line.
In the case of the vortex half--ring the intersection plane
$\Sigma$ has to include the center of the semicircle
formed by the VL and the axis $x$
(see fig.~\ref{halfring}).
The specific feature of magnetic structures in the form
of a vortex half--ring is self--consistent localization
of both the magnetic field and supercurrent in the vicinity
of the VL.
Since the magnetic field and the supercurrent vanish
at large distances from the VL, the total magnetic flux
$\Phi_{\Sigma}$ across the plane $\Sigma$ is equal exactly to zero:
\beq
\Phi_{\Sigma} = \Phi_{\Sigma_{\ds -}} +
                \Phi_{\Sigma_{\ds +}} \equiv 0,         \label{TotalFlux}
\eeq
Here $\Phi_{\Sigma_{\ds \pm}}$ are the magnetic fluxes across the
corresponding half--planes $\Sigma_\pm$.
Using the Gauss theorem, it is easy to verify that the fluxes
$\Phi_{\Sigma_{\ds \pm}}$ do not depend on the orientation of the
intersection plane $\Sigma$ relative to the surface of the
superconductor.
So, by choosing for simplicity the plane
$y=0$ as the cross--section $\Sigma$,
one can obtain:
\beq
\Phi_{\Sigma_{\ds -}} = \Phi_v + \Phi_d,                \label{MinusFlux}
\eeq
$$
\Phi_v = \int\limits_{-\infty}^{+\infty} dx %
         \int\limits_{-\infty}^0 dz\, H_v(z,x), \qquad
\Phi_d = \int\limits_{-\infty}^{+\infty} dx %
         \int\limits_{-\infty}^0 dz\, H_d^y(x,0,z),
$$
\beq
\Phi_{\Sigma_{\ds +}} \equiv \Phi_s =
         \int\limits_{-\infty}^{+\infty} dx %
         \int\limits_0^{+\infty} dz H_s^y(x,0,z).       \label{PlusFlux}
\eeq
The term $\Phi_v$ is the magnetic flux in the solitary
toroidal Abrikosov vortex and was calculated in
\cite{Kozlov93}
\beq
\Phi_v(\rho_v) = 1 - \rho_v K_1( \rho_v),               \label{TorusFlux} %
\eeq
and the magnetic fluxes $\Phi_d$ and $\Phi_s$ are determined completely
by the spectral function  $S_v(q,u)$:
\beq
\Phi_d = -\frac{2}{\pi} \int\limits_0^{+\infty} du\, %
         \frac{\ds S_v(0,u)}
              {\ds u+\sqrt{1+u^2}},                     \label{dFlux} %
\eeq
\beq
\Phi_s = -\frac{2}{\pi} \int\limits_0^{+\infty} du\, %
     \frac{\ds \sqrt{1+u^2}\; S_v(0,u)}
          {\ds u \left(u+\sqrt{1+u^2} \right)}.         \label{sFlux} %
\eeq
Using the relations (\ref{TorusFlux},\ref{dFlux},\ref{sFlux})
and the definition of the spectral function $S_v(q,u)$
(\ref{FourSpec}) it is easy to verify that the total magnetic
flux $\Phi_{\Sigma}$ is really equal to zero.
The results of the simulations of the magnetic fluxes $\Phi_v$
and $\Phi_{\Sigma_{\ds -}}$ versus the vortex radius $\rho_v$
are shown in fig.~\ref{flux}.
The magnetic flux $\Phi_{\Sigma_{\ds -}}$ of the vortex half--ring
depends strongly on the VL radius $\rho_v$ and tends asymptotically to
the quantum of magnetic flux $\Phi_0$ at large values of $\rho_v$.
The flux $\Phi_{\Sigma_{\ds -}}$ differs noticeably from the
flux of a toroidal Abrikosov vortex $\Phi_v$ .
It means that the surface of a superconductor affects significantly
the magnetic field structure of the vortex half--ring, thus largely
reducing the value of the magnetic field in the vortex core area
\cite{Fritz93}.

In the region $z \le 0$, occupied by the superconductor, the London
equation (\ref{eqnLondon}) implies the following expression for
the free energy $G_-$
\cite{Abrikosov&Fund}:
\beq
G_- = \frac{1}{8 \pi} %
\int\limits_{\ds z=0} dx\, dy {\Bigl[ {\bf H} \times \rot{\bf H} \Bigr]}_z %
+ \frac{1}{8 \pi} \int\limits_{\ds z \le 0} dV {\bf H} %
\Bigl( {\bf H}+ \rot\rot {\bf H} \Bigr).               \label{enLondon} %
\eeq
The energy contribution of the magnetic field ${\bf H}_s$
outside the superconductor $(z>0)$ is defined in the usual way:
\beq
G_+ = \frac{1}{8 \pi} %
\int\limits_{\ds z > 0} dV {\left( {\bf H_s} \right) }^2.\label{enField} %
\eeq
The total free energy $G_0$ is the sum of the expressions
(\ref{enLondon},\ref{enField}):
\beq
G_0=G_{-} + G_{+} .                                       \label{enFree0} %
\eeq
To find the free energy of the vortex half--ring it is necessary
to substitute the solutions
(\ref{Superpos},\ref{solLondonR},\ref{solLondon0},\ref{defUs},%
\ref{solLaplace},\ref{solC},\ref{SpecFunc}) %
describing the magnetic field distributions ${\bf H}$ and ${\bf H}_s$
in the expressions (\ref{enLondon},\ref{enField},\ref{enFree0}).
It is convenient to extract the contribution of the superconductor
surface and represent the energy $G_0$ as the sum:
\beq
G_0 = G_v + G_s.                                          \label{enFree} %
\eeq
Here the first term $G_v$ is the free energy of the half of
the toroidal Abrikosov vortex. The magnitude of $G_v$ is fully
determined by the value of the magnetic field in the center
of the vortex line
\cite{Kozlov91,Kozlov93}:
\beq
G_v = \frac{\rho_v}{8} H_v(\rho_v,0).                   \label{enTorus} %
\eeq
The second term $G_s$ includes the total influence of the surface $S$
and is expressed in terms of the spectral function $S_v(q,u)$:
\beqr
& &G_s = \frac{\rho_v}{\ds 2 \pi^3} %
\int\limits_0^{+\infty} dq \int\limits_0^{+\infty} du %
\frac{\ds u \sqrt{1+q^2+u^2}\; S_v(q,u)\; D(q,u) } %
     { \ds \sqrt{q^2+u^2}\; %
       \left( \sqrt{1+q^2+u^2}
                +\sqrt{q^2+u^2} \right) },              \label{enSurface} \\
& &D(q,u) = \frac{\ds \sin{(u \rho_v)}}{u \rho_v} %
-\int\limits_0^{\pi/2} d\varphi %
\Bigl[ \cos{\varphi} \cos{(u \rho_v \sin{\varphi})} \nonumber \\ %
      & &\qquad + \frac{\ds q^2+u^2}{u \sqrt{1+q^2+u^2}} %
                \sin{\varphi} \sin{(u \rho_v \sin{\varphi})} %
\Bigl] \exp{(\ds -\rho_v \cos{\varphi} \sqrt{1+q^2+u^2})}.\nonumber
\eeqr
Expression (\ref{solLondonR}) has a logarithmic divergence at the VL
center ($\rho=\rho_v$, $x=0$), since the London equation
(\ref{eqnLondonR}) is not valid in the region of the normal vortex
core.
To compute the energy $G_v$ (\ref{enTorus}), the magnetic field
$H_v(\rho_v,0)$ was thus truncated at the distance $\xi$ from the VL
center.
The details of the numerical simulations of the magnetic field
$H_v(\rho_v,0)$ are presented in Appendix C.
Figure~\ref{Freen} gives the dependencies of the energies $G_v$ and $G_s$
on the radius of the vortex half--ring $\rho_v$.
Let us note that $G_s \ll G_v$ at any values of the vortex radius.
It means that the free energy $G_0$ is modified insignificantly by
the surface of the superconductor and the estimate
\beq
G_0(\rho_v) \simeq G_v(\rho_v)                            \label{enFreeAs} %
\eeq
accurately approximates the exact value of the free energy
of the vortex half--ring.

Thus, the free energy $G_0(\rho_v)$ of the Abrikosov vortex
in the form of a semicircle of radius $\rho_v$
exceeds very slightly the energy $G_v(\rho_v)$ of
half of the toroidal vortex.
At the same time, the magnetic field structures of these
vortices differ noticeably, especially inside the layer
with thickness $\sim \lambda$ near the surface.
The surface changes the structure of the toroidal vortex,
such that the value of the magnetic field in the core
of the vortex decreases.
As a consequence, the magnetic fluxes $\Phi_{\Sigma_{\ds \pm}}$
(\ref{MinusFlux},\ref{PlusFlux}) of the vortex half--ring
are smaller than the magnetic flux $\Phi_v$ (\ref{TorusFlux})
of the toroidal Abrikosov vortex even if the size of
the vortices exceeds the magnetic field penetration depth
$\lambda$ ($\rho_v \gtrsim 1$).
Due to the continuity of the magnetic flux, the flux line
creates the scattering field ${\bf H}_s$ outside the
superconductor.
The energy of the magnetic field ${\bf H}_s$ makes an
important contribution to the total free energy of the
vortex half--ring and neutralizes the
decrease of the London energy $G_{-}$ (\ref{enLondon}),
forced by the distortions of the vortex structure, induced by
the surface of superconductor.
%
%
%
%%%%%%%%%%%%%%%%%%%%%%%%%%%%%%%%%%%%%%%%%%%%%%%%%%%%%%%%%%%%%%%%%%%%%%
%            3. VORTEX HALF-RING IN AN APPLIED FIELD                 %
%%%%%%%%%%%%%%%%%%%%%%%%%%%%%%%%%%%%%%%%%%%%%%%%%%%%%%%%%%%%%%%%%%%%%%
\vspace{10mm}
\begin{center}
{\small\bf 3. VORTEX HALF-RING IN AN APPLIED MAGNETIC FIELD}
\end{center}

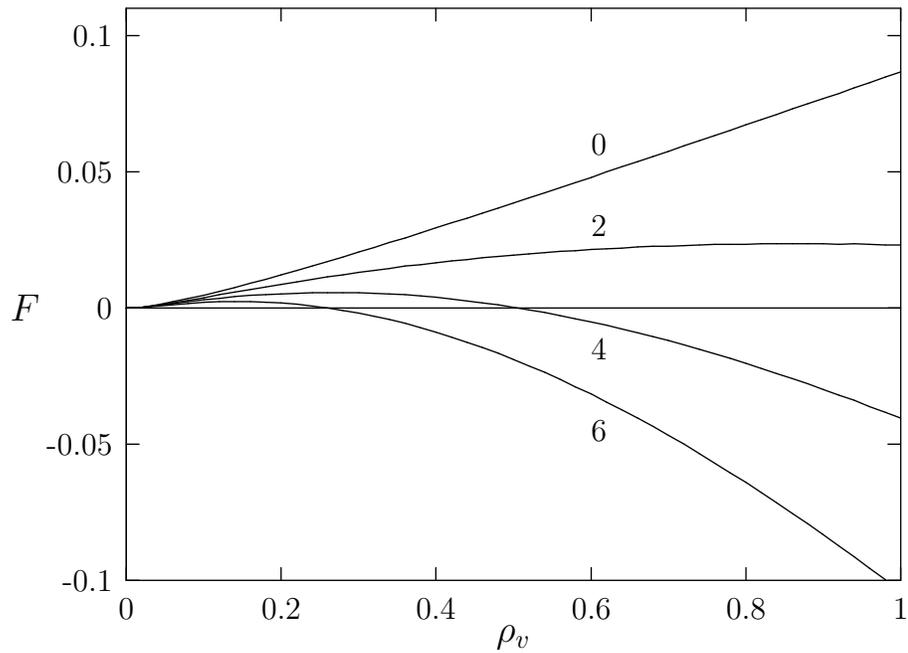
\begin{figure}[htb]
%\input{gibbs.pic}
% GNUPLOT: LaTeX picture with emtex specials
\setlength{\unitlength}{0.240900pt}
\ifx\plotpoint\undefined\newsavebox{\plotpoint}\fi
\sbox{\plotpoint}{\rule[-0.500pt]{1.000pt}{1.000pt}}%
\special{em:linewidth 0.50pt}%
\begin{picture}(1500,1080)(0,0)
\font\gnuplot=cmr12 at 12pt
\gnuplot
\put(220,541){\special{em:moveto}}
\put(1436,541){\special{em:lineto}}
\put(220,113){\special{em:moveto}}
\put(220,1012){\special{em:lineto}}
\put(220,113){\special{em:moveto}}
\put(240,113){\special{em:lineto}}
\put(1436,113){\special{em:moveto}}
\put(1416,113){\special{em:lineto}}
\put(198,113){\makebox(0,0)[r]{-0.1}}
\put(220,327){\special{em:moveto}}
\put(240,327){\special{em:lineto}}
\put(1436,327){\special{em:moveto}}
\put(1416,327){\special{em:lineto}}
\put(198,327){\makebox(0,0)[r]{-0.05}}
\put(220,541){\special{em:moveto}}
\put(240,541){\special{em:lineto}}
\put(1436,541){\special{em:moveto}}
\put(1416,541){\special{em:lineto}}
\put(198,541){\makebox(0,0)[r]{0}}
\put(220,755){\special{em:moveto}}
\put(240,755){\special{em:lineto}}
\put(1436,755){\special{em:moveto}}
\put(1416,755){\special{em:lineto}}
\put(198,755){\makebox(0,0)[r]{0.05}}
\put(220,969){\special{em:moveto}}
\put(240,969){\special{em:lineto}}
\put(1436,969){\special{em:moveto}}
\put(1416,969){\special{em:lineto}}
\put(198,969){\makebox(0,0)[r]{0.1}}
\put(220,113){\special{em:moveto}}
\put(220,133){\special{em:lineto}}
\put(220,1012){\special{em:moveto}}
\put(220,992){\special{em:lineto}}
\put(220,68){\makebox(0,0){0}}
\put(463,113){\special{em:moveto}}
\put(463,133){\special{em:lineto}}
\put(463,1012){\special{em:moveto}}
\put(463,992){\special{em:lineto}}
\put(463,68){\makebox(0,0){0.2}}
\put(706,113){\special{em:moveto}}
\put(706,133){\special{em:lineto}}
\put(706,1012){\special{em:moveto}}
\put(706,992){\special{em:lineto}}
\put(706,68){\makebox(0,0){0.4}}
\put(950,113){\special{em:moveto}}
\put(950,133){\special{em:lineto}}
\put(950,1012){\special{em:moveto}}
\put(950,992){\special{em:lineto}}
\put(950,68){\makebox(0,0){0.6}}
\put(1193,113){\special{em:moveto}}
\put(1193,133){\special{em:lineto}}
\put(1193,1012){\special{em:moveto}}
\put(1193,992){\special{em:lineto}}
\put(1193,68){\makebox(0,0){0.8}}
\put(1436,113){\special{em:moveto}}
\put(1436,133){\special{em:lineto}}
\put(1436,1012){\special{em:moveto}}
\put(1436,992){\special{em:lineto}}
\put(1436,68){\makebox(0,0){1}}
\put(220,113){\special{em:moveto}}
\put(1436,113){\special{em:lineto}}
\put(1436,1012){\special{em:lineto}}
\put(220,1012){\special{em:lineto}}
\put(220,113){\special{em:lineto}}
\put(45,562){\makebox(0,0){   }}
\put(828,23){\makebox(0,0){$\rho_v$}}
\put(828,1057){\makebox(0,0){     }}
\put(38,541){\makebox(0,0)[l]{$F$}}
\put(950,798){\makebox(0,0)[l]{0}}
\put(950,670){\makebox(0,0)[l]{2}}
\put(950,348){\makebox(0,0)[l]{6}}
\put(950,477){\makebox(0,0)[l]{4}}
\put(220,541){\special{em:moveto}}
\put(244,542){\special{em:lineto}}
\put(269,546){\special{em:lineto}}
\put(293,551){\special{em:lineto}}
\put(317,556){\special{em:lineto}}
\put(342,561){\special{em:lineto}}
\put(366,567){\special{em:lineto}}
\put(390,573){\special{em:lineto}}
\put(415,580){\special{em:lineto}}
\put(439,586){\special{em:lineto}}
\put(463,593){\special{em:lineto}}
\put(488,600){\special{em:lineto}}
\put(512,607){\special{em:lineto}}
\put(536,614){\special{em:lineto}}
\put(560,621){\special{em:lineto}}
\put(585,629){\special{em:lineto}}
\put(609,636){\special{em:lineto}}
\put(633,644){\special{em:lineto}}
\put(658,651){\special{em:lineto}}
\put(682,659){\special{em:lineto}}
\put(706,667){\special{em:lineto}}
\put(731,675){\special{em:lineto}}
\put(755,682){\special{em:lineto}}
\put(779,690){\special{em:lineto}}
\put(804,698){\special{em:lineto}}
\put(828,706){\special{em:lineto}}
\put(852,714){\special{em:lineto}}
\put(877,722){\special{em:lineto}}
\put(901,730){\special{em:lineto}}
\put(925,738){\special{em:lineto}}
\put(950,746){\special{em:lineto}}
\put(974,755){\special{em:lineto}}
\put(998,763){\special{em:lineto}}
\put(1023,771){\special{em:lineto}}
\put(1047,779){\special{em:lineto}}
\put(1071,787){\special{em:lineto}}
\put(1096,796){\special{em:lineto}}
\put(1120,804){\special{em:lineto}}
\put(1144,812){\special{em:lineto}}
\put(1168,820){\special{em:lineto}}
\put(1193,829){\special{em:lineto}}
\put(1217,837){\special{em:lineto}}
\put(1241,845){\special{em:lineto}}
\put(1266,854){\special{em:lineto}}
\put(1290,862){\special{em:lineto}}
\put(1314,870){\special{em:lineto}}
\put(1339,878){\special{em:lineto}}
\put(1363,887){\special{em:lineto}}
\put(1387,895){\special{em:lineto}}
\put(1412,904){\special{em:lineto}}
\put(1436,912){\special{em:lineto}}
\put(220,541){\special{em:moveto}}
\put(244,542){\special{em:lineto}}
\put(269,545){\special{em:lineto}}
\put(293,549){\special{em:lineto}}
\put(317,553){\special{em:lineto}}
\put(342,557){\special{em:lineto}}
\put(366,562){\special{em:lineto}}
\put(390,566){\special{em:lineto}}
\put(415,570){\special{em:lineto}}
\put(439,574){\special{em:lineto}}
\put(463,578){\special{em:lineto}}
\put(488,582){\special{em:lineto}}
\put(512,586){\special{em:lineto}}
\put(536,590){\special{em:lineto}}
\put(560,593){\special{em:lineto}}
\put(585,597){\special{em:lineto}}
\put(609,600){\special{em:lineto}}
\put(633,603){\special{em:lineto}}
\put(658,607){\special{em:lineto}}
\put(682,609){\special{em:lineto}}
\put(706,612){\special{em:lineto}}
\put(731,615){\special{em:lineto}}
\put(755,617){\special{em:lineto}}
\put(779,620){\special{em:lineto}}
\put(804,622){\special{em:lineto}}
\put(828,624){\special{em:lineto}}
\put(852,626){\special{em:lineto}}
\put(877,628){\special{em:lineto}}
\put(901,630){\special{em:lineto}}
\put(925,631){\special{em:lineto}}
\put(950,633){\special{em:lineto}}
\put(974,634){\special{em:lineto}}
\put(998,635){\special{em:lineto}}
\put(1023,637){\special{em:lineto}}
\put(1047,638){\special{em:lineto}}
\put(1071,638){\special{em:lineto}}
\put(1096,639){\special{em:lineto}}
\put(1120,640){\special{em:lineto}}
\put(1144,641){\special{em:lineto}}
\put(1168,641){\special{em:lineto}}
\put(1193,641){\special{em:lineto}}
\put(1217,642){\special{em:lineto}}
\put(1241,642){\special{em:lineto}}
\put(1266,642){\special{em:lineto}}
\put(1290,642){\special{em:lineto}}
\put(1314,642){\special{em:lineto}}
\put(1339,641){\special{em:lineto}}
\put(1363,642){\special{em:lineto}}
\put(1387,641){\special{em:lineto}}
\put(1412,640){\special{em:lineto}}
\put(1436,640){\special{em:lineto}}
\put(220,541){\special{em:moveto}}
\put(244,542){\special{em:lineto}}
\put(269,545){\special{em:lineto}}
\put(293,548){\special{em:lineto}}
\put(317,551){\special{em:lineto}}
\put(342,554){\special{em:lineto}}
\put(366,556){\special{em:lineto}}
\put(390,559){\special{em:lineto}}
\put(415,561){\special{em:lineto}}
\put(439,562){\special{em:lineto}}
\put(463,563){\special{em:lineto}}
\put(488,564){\special{em:lineto}}
\put(512,565){\special{em:lineto}}
\put(536,565){\special{em:lineto}}
\put(560,565){\special{em:lineto}}
\put(585,565){\special{em:lineto}}
\put(609,564){\special{em:lineto}}
\put(633,563){\special{em:lineto}}
\put(658,562){\special{em:lineto}}
\put(682,560){\special{em:lineto}}
\put(706,558){\special{em:lineto}}
\put(731,555){\special{em:lineto}}
\put(755,552){\special{em:lineto}}
\put(779,549){\special{em:lineto}}
\put(804,546){\special{em:lineto}}
\put(828,542){\special{em:lineto}}
\put(852,538){\special{em:lineto}}
\put(877,534){\special{em:lineto}}
\put(901,529){\special{em:lineto}}
\put(925,524){\special{em:lineto}}
\put(950,519){\special{em:lineto}}
\put(974,514){\special{em:lineto}}
\put(998,508){\special{em:lineto}}
\put(1023,502){\special{em:lineto}}
\put(1047,496){\special{em:lineto}}
\put(1071,490){\special{em:lineto}}
\put(1096,483){\special{em:lineto}}
\put(1120,476){\special{em:lineto}}
\put(1144,469){\special{em:lineto}}
\put(1168,462){\special{em:lineto}}
\put(1193,454){\special{em:lineto}}
\put(1217,446){\special{em:lineto}}
\put(1241,438){\special{em:lineto}}
\put(1266,430){\special{em:lineto}}
\put(1290,422){\special{em:lineto}}
\put(1314,413){\special{em:lineto}}
\put(1339,404){\special{em:lineto}}
\put(1363,396){\special{em:lineto}}
\put(1387,386){\special{em:lineto}}
\put(1412,377){\special{em:lineto}}
\put(1436,368){\special{em:lineto}}
\put(220,541){\special{em:moveto}}
\put(244,542){\special{em:lineto}}
\put(269,544){\special{em:lineto}}
\put(293,546){\special{em:lineto}}
\put(317,548){\special{em:lineto}}
\put(342,550){\special{em:lineto}}
\put(366,551){\special{em:lineto}}
\put(390,551){\special{em:lineto}}
\put(415,551){\special{em:lineto}}
\put(439,550){\special{em:lineto}}
\put(463,549){\special{em:lineto}}
\put(488,547){\special{em:lineto}}
\put(512,544){\special{em:lineto}}
\put(536,541){\special{em:lineto}}
\put(560,537){\special{em:lineto}}
\put(585,533){\special{em:lineto}}
\put(609,528){\special{em:lineto}}
\put(633,523){\special{em:lineto}}
\put(658,517){\special{em:lineto}}
\put(682,510){\special{em:lineto}}
\put(706,503){\special{em:lineto}}
\put(731,495){\special{em:lineto}}
\put(755,487){\special{em:lineto}}
\put(779,479){\special{em:lineto}}
\put(804,470){\special{em:lineto}}
\put(828,460){\special{em:lineto}}
\put(852,450){\special{em:lineto}}
\put(877,440){\special{em:lineto}}
\put(901,429){\special{em:lineto}}
\put(925,417){\special{em:lineto}}
\put(950,406){\special{em:lineto}}
\put(974,393){\special{em:lineto}}
\put(998,381){\special{em:lineto}}
\put(1023,368){\special{em:lineto}}
\put(1047,355){\special{em:lineto}}
\put(1071,341){\special{em:lineto}}
\put(1096,327){\special{em:lineto}}
\put(1120,312){\special{em:lineto}}
\put(1144,297){\special{em:lineto}}
\put(1168,282){\special{em:lineto}}
\put(1193,267){\special{em:lineto}}
\put(1217,251){\special{em:lineto}}
\put(1241,235){\special{em:lineto}}
\put(1266,218){\special{em:lineto}}
\put(1290,202){\special{em:lineto}}
\put(1314,185){\special{em:lineto}}
\put(1339,167){\special{em:lineto}}
\put(1363,150){\special{em:lineto}}
\put(1387,132){\special{em:lineto}}
\put(1412,114){\special{em:lineto}}
\put(1413,113){\special{em:lineto}}
\end{picture}
\caption{Gibbs free energy $F$ of the vortex half--ring vs.
the radius of the vortex line $\rho_v$ for several values of
an applied magnetic field $H_0$ ($\kappa=100$). %
The numbers near the curves denote the value of the field %
$H_0$ in the units of the lower critical field $H_{c1}$. } %
\label{Gibbsen}
\end{figure}

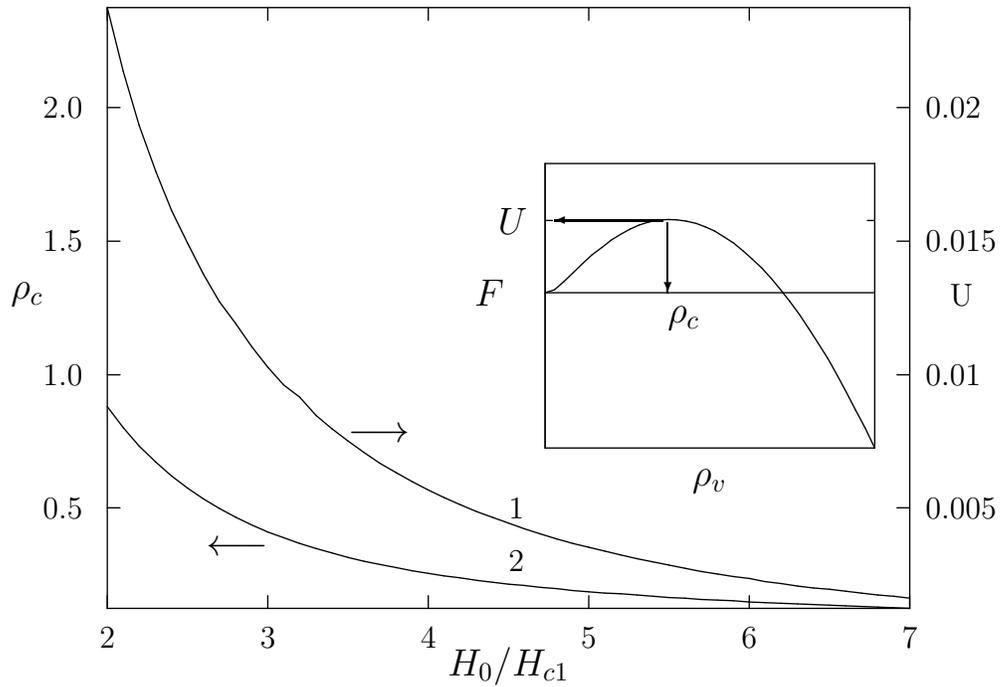
\begin{figure}[htb]
%\input{barrier.pic}
% GNUPLOT: LaTeX picture with emtex specials
\setlength{\unitlength}{0.240900pt}
\ifx\plotpoint\undefined\newsavebox{\plotpoint}\fi
\sbox{\plotpoint}{\rule[-0.500pt]{1.000pt}{1.000pt}}%
\special{em:linewidth 0.5pt}%
\begin{picture}(1500,1080)(0,0)
\font\gnuplot=cmr12 at 12pt
\gnuplot
\put(176,271){\special{em:moveto}}
\put(196,271){\special{em:lineto}}
\put(1436,271){\special{em:moveto}}
\put(1416,271){\special{em:lineto}}
\put(154,271){\makebox(0,0)[r]{ }}
\put(176,480){\special{em:moveto}}
\put(196,480){\special{em:lineto}}
\put(1436,480){\special{em:moveto}}
\put(1416,480){\special{em:lineto}}
\put(154,480){\makebox(0,0)[r]{ }}
\put(176,690){\special{em:moveto}}
\put(196,690){\special{em:lineto}}
\put(1436,690){\special{em:moveto}}
\put(1416,690){\special{em:lineto}}
\put(154,690){\makebox(0,0)[r]{ }}
\put(176,900){\special{em:moveto}}
\put(196,900){\special{em:lineto}}
\put(1436,900){\special{em:moveto}}
\put(1416,900){\special{em:lineto}}
\put(154,900){\makebox(0,0)[r]{ }}
\put(176,113){\special{em:moveto}}
\put(176,133){\special{em:lineto}}
\put(176,1057){\special{em:moveto}}
\put(176,1037){\special{em:lineto}}
\put(176,68){\makebox(0,0){2}}
\put(428,113){\special{em:moveto}}
\put(428,133){\special{em:lineto}}
\put(428,1057){\special{em:moveto}}
\put(428,1037){\special{em:lineto}}
\put(428,68){\makebox(0,0){3}}
\put(680,113){\special{em:moveto}}
\put(680,133){\special{em:lineto}}
\put(680,1057){\special{em:moveto}}
\put(680,1037){\special{em:lineto}}
\put(680,68){\makebox(0,0){4}}
\put(932,113){\special{em:moveto}}
\put(932,133){\special{em:lineto}}
\put(932,1057){\special{em:moveto}}
\put(932,1037){\special{em:lineto}}
\put(932,68){\makebox(0,0){5}}
\put(1184,113){\special{em:moveto}}
\put(1184,133){\special{em:lineto}}
\put(1184,1057){\special{em:moveto}}
\put(1184,1037){\special{em:lineto}}
\put(1184,68){\makebox(0,0){6}}
\put(1436,113){\special{em:moveto}}
\put(1436,133){\special{em:lineto}}
\put(1436,1057){\special{em:moveto}}
\put(1436,1037){\special{em:lineto}}
\put(1436,68){\makebox(0,0){7}}
\put(176,113){\special{em:moveto}}
\put(1436,113){\special{em:lineto}}
\put(1436,1057){\special{em:lineto}}
\put(176,1057){\special{em:lineto}}
\put(176,113){\special{em:lineto}}
\put(806,23){\makebox(0,0){$H_0/H_{c1}$}}
\put(25,606){\makebox(0,0)[l]{$\rho_c$}}
\put(806,187){\makebox(0,0)[l]{2}}
\put(806,271){\makebox(0,0)[l]{1}}
\put(1499,606){\makebox(0,0)[l]{U}}
\put(75,271){\makebox(0,0)[l]{0.5}}
\put(75,480){\makebox(0,0)[l]{1.0}}
\put(75,690){\makebox(0,0)[l]{1.5}}
\put(75,900){\makebox(0,0)[l]{2.0}}
\put(1461,271){\makebox(0,0)[l]{0.005}}
\put(1461,480){\makebox(0,0)[l]{0.01}}
\put(1461,690){\makebox(0,0)[l]{0.015}}
\put(1461,900){\makebox(0,0)[l]{0.02}}
\put(554,386){\makebox(0,0)[l]{$\longrightarrow$}}
\put(428,208){\makebox(0,0)[r]{$\longleftarrow$}}
\put(176,1057){\special{em:moveto}}
\put(201,957){\special{em:lineto}}
\put(226,872){\special{em:lineto}}
\put(252,800){\special{em:lineto}}
\put(277,738){\special{em:lineto}}
\put(302,687){\special{em:lineto}}
\put(327,638){\special{em:lineto}}
\put(352,595){\special{em:lineto}}
\put(378,560){\special{em:lineto}}
\put(403,524){\special{em:lineto}}
\put(428,492){\special{em:lineto}}
\put(453,464){\special{em:lineto}}
\put(478,445){\special{em:lineto}}
\put(504,416){\special{em:lineto}}
\put(529,395){\special{em:lineto}}
\put(554,376){\special{em:lineto}}
\put(579,358){\special{em:lineto}}
\put(604,341){\special{em:lineto}}
\put(630,326){\special{em:lineto}}
\put(655,312){\special{em:lineto}}
\put(680,299){\special{em:lineto}}
\put(705,287){\special{em:lineto}}
\put(730,276){\special{em:lineto}}
\put(756,265){\special{em:lineto}}
\put(781,256){\special{em:lineto}}
\put(806,247){\special{em:lineto}}
\put(831,238){\special{em:lineto}}
\put(856,230){\special{em:lineto}}
\put(882,222){\special{em:lineto}}
\put(907,215){\special{em:lineto}}
\put(932,209){\special{em:lineto}}
\put(957,203){\special{em:lineto}}
\put(982,197){\special{em:lineto}}
\put(1008,191){\special{em:lineto}}
\put(1033,186){\special{em:lineto}}
\put(1058,181){\special{em:lineto}}
\put(1083,176){\special{em:lineto}}
\put(1108,171){\special{em:lineto}}
\put(1134,167){\special{em:lineto}}
\put(1159,163){\special{em:lineto}}
\put(1184,160){\special{em:lineto}}
\put(1209,155){\special{em:lineto}}
\put(1234,152){\special{em:lineto}}
\put(1260,148){\special{em:lineto}}
\put(1285,145){\special{em:lineto}}
\put(1310,143){\special{em:lineto}}
\put(1335,140){\special{em:lineto}}
\put(1360,137){\special{em:lineto}}
\put(1386,134){\special{em:lineto}}
\put(1411,132){\special{em:lineto}}
\put(1436,129){\special{em:lineto}}
\put(176,430){\special{em:moveto}}
\put(201,397){\special{em:lineto}}
\put(226,368){\special{em:lineto}}
\put(252,343){\special{em:lineto}}
\put(277,321){\special{em:lineto}}
\put(302,302){\special{em:lineto}}
\put(327,285){\special{em:lineto}}
\put(352,270){\special{em:lineto}}
\put(378,256){\special{em:lineto}}
\put(403,244){\special{em:lineto}}
\put(428,233){\special{em:lineto}}
\put(453,224){\special{em:lineto}}
\put(478,215){\special{em:lineto}}
\put(504,207){\special{em:lineto}}
\put(529,200){\special{em:lineto}}
\put(554,193){\special{em:lineto}}
\put(579,187){\special{em:lineto}}
\put(604,182){\special{em:lineto}}
\put(630,177){\special{em:lineto}}
\put(655,172){\special{em:lineto}}
\put(680,168){\special{em:lineto}}
\put(705,164){\special{em:lineto}}
\put(730,161){\special{em:lineto}}
\put(756,157){\special{em:lineto}}
\put(781,154){\special{em:lineto}}
\put(806,151){\special{em:lineto}}
\put(831,149){\special{em:lineto}}
\put(856,146){\special{em:lineto}}
\put(882,144){\special{em:lineto}}
\put(907,141){\special{em:lineto}}
\put(932,139){\special{em:lineto}}
\put(957,137){\special{em:lineto}}
\put(982,136){\special{em:lineto}}
\put(1008,134){\special{em:lineto}}
\put(1033,132){\special{em:lineto}}
\put(1058,130){\special{em:lineto}}
\put(1083,129){\special{em:lineto}}
\put(1108,127){\special{em:lineto}}
\put(1134,126){\special{em:lineto}}
\put(1159,125){\special{em:lineto}}
\put(1184,123){\special{em:lineto}}
\put(1209,122){\special{em:lineto}}
\put(1234,121){\special{em:lineto}}
\put(1260,120){\special{em:lineto}}
\put(1285,119){\special{em:lineto}}
\put(1310,118){\special{em:lineto}}
\put(1335,117){\special{em:lineto}}
\put(1360,116){\special{em:lineto}}
\put(1386,115){\special{em:lineto}}
\put(1411,114){\special{em:lineto}}
\put(1436,113){\special{em:lineto}}
\end{picture}
\caption{Surface barrier $U$ (1) and critical radius of the vortex
half--ring $\rho_c$ (2) vs. the value of the applied magnetic
field $H_0$ ($\kappa=100$). %
The inset sketches the part of the curve $F(\rho_v)$.
The Gibbs free energy $F(\rho_v)$ has its maximum $U$ at
the critical radius $\rho_c$: $F(\rho_c)=U$. } %
\label{Barrier}
\end{figure}

The behavior of a superconductor in an applied magnetic field
${\bf H}_0$ is controlled by the Gibbs free energy
\cite{Abrikosov&Fund}:
\beq
F = G - \frac{1}{4 \pi}
\int\limits_{z \le 0} dV
\left( {\bf H} {\bf H}_0 \right).                       \label{enGibbs} %
\eeq
If the applied field ${\bf H}_0$ is directed along the
surface $S$, the magnetic field distribution in the superconductor
half--space $(z \le 0)$ (see fig.~\ref{halfring}) can be
represented as follows:
\beq
{\bf H} = {\bf H}_V + {\bf H}_m.                        \label{Superpos1} %
\eeq
The first term ${\bf H}_V$ obeys equation (\ref{eqnLondon})
with the boundary conditions (\ref{BoundCond}) and describes the
magnetic field structure of the Abrikosov vortex with
arbitrary vortex line, if the applied field is absent.
The second term ${\bf H}_m = {\bf H}_0 \exp{(z)}$ is the Meissner
(vortex--free) distribution of the magnetic field near the surface
of the superconductor.
The free energy $G$ is defined by the expressions
(\ref{enLondon},\ref{enField},\ref{enFree0}),
where the magnetic field ${\bf H}$ in the superconductor
satisfies relation (\ref{Superpos1}).
Taking into account that the field ${\bf H}_m$ on the
superconductor surface $S$ coincides with the
applied magnetic field ${\bf H}_0$, the Gibbs free energy
$F$ (\ref{enGibbs}) measured relative to the free energy
of the Meissner state can be expressed as:
\beq
F = G_V +
\frac{1}{4 \pi}\int\limits_{l_v} %
d{\bf l}_v {\bf H}_m - %
\frac{1}{4 \pi}\int\limits_{l_v} %
d{\bf l}_v {\bf H}_0 .                                  \label{enGibbs1} %
\eeq
Here $G_V$ is the free energy of an Abrikosov vortex
described by the term ${\bf H}_V$ in (\ref{Superpos1})
in a zero applied magnetic field.
The VL location of this vortex in space
is determined by arbitrary curve $l_v$
and the vector $d{\bf l}_v=d l_v {\bf e}_v$ is directed along
the vortex line.
It should be noted that expression (\ref{enGibbs1}) sets
the relation between the free energy of the Abrikosov vortex
$G_V$ and the Gibbs free energy $F$ for arbitrary shape
of the vortex line.
Using (\ref{enGibbs1}), it is easy to write
the Gibbs free energy of the Abrikosov vortex in the form of
a semicircle of arbitrary radius $\rho_v$:
\beq
F(\rho_v) = G_0(\rho_v)-\frac{H_0 \rho_v}{2 \pi}\; %
\Bigl[ 1-\int\limits_0^{\pi/2} d\varphi \sin{\varphi} %
        \exp{(-\rho_v \sin\varphi)} \Bigr].             \label{enGibbsEx} %
\eeq
Figure~\ref{Gibbsen} gives the dependencies of the Gibbs free
energy $F$ (\ref{enGibbsEx}) on the radius of the vortex
half--ring $\rho_v$ for some values of an applied
magnetic field $H_0$.
The type of the curve $F(\rho_v)$ depends strongly on the
value of the applied magnetic field $H_0$.
Taking into account the equality (\ref{enFreeAs}) and
considering that in the case $\rho_v \gg 1$ the following
approximate expression for the free energy $G_v(\rho_v)$
holds
\beq
G_v(\rho_v) \simeq \frac{H_{c1}}{4} \rho_v,
\qquad \rho_v \gg 1,                                    \label{enFreeL} %
\eeq
one can obtain the asymptotic behavior of the Gibbs free energy
$F(\rho_v)$ for $\rho_v \gg 1$:
\beq
F(\rho_v) \simeq \frac{\rho_v}{2 \pi}
\left[ \frac{\pi}{2} H_{c1} - H_0 \right],
\qquad \rho_v \gg 1.                                    \label{enGibbsL} %
\eeq
Here $H_{c1} = \ln{(\kappa)} / 4 \pi$ is the lower critical magnetic
field for the London model.
If $H_0 < H_{c1}^* = \pi H_{c1} /2$, the Gibbs free energy $F(\rho_v)$
is positive for any $\rho_v \neq 0$, and the Meissner state has
the minimal energy.
So, the existence of vortex half--rings in the superconductor is forbidden,
and only vortex--free distributions of the magnetic field and
supercurrent are possible.
In the case of $H_0 = H_{c1}^*$, the Gibbs free energy of a
state containing a vortex half--ring of a large radius
$\rho_v \gg 1$ matches the energy of the Meissner state.
It should be noted that the vortex half--ring is
asymptotically stable in this case, because
$$
\lim_{\rho_v\to\infty} \frac{dF}{d\rho_v} = 0.
$$
So, the field $H_{c1}^*$ is the minimal applied magnetic field, when
it is energetically favorable for a vortex half--ring to be formed
near the flat surface of type--II superconductor.
If $H_0 > H_{c1}^*$, there is a critical radius $\rho_c$ of the
vortex half--ring at which the curve $F(\rho_v)$ has its maximum $U$
(see the inset in fig.~\ref{Barrier}).
Figure~\ref{Barrier} gives the dependencies of the height of
the surface energy barrier $U$ and the critical radius $\rho_c$
on the value of an applied magnetic field $H_0$.
Since the Gibbs free energy $F(\rho_v)$ decreases monotonically
at $\rho_v > \rho_c$, the energy barrier $U$ prevents nucleation
of expanding vortex loops near the surface.
The barrier height $U$ is reduced by increasing
the applied field $H_0$.
When the applied field $H_0$ is large enough for the critical radius
$\rho_c$ to become of the order of the coherence length $\xi$,
the surface barrier $U$ vanishes.
Using the approximate expression for the free energy $G_v$
of the vortex half--ring
\cite{Kolomeisky91,Chen93}
\beq
G_v(\rho_v) \simeq \frac{\rho_v}{4}
\left[ H_{c1} + \frac{\ln(\rho_v)}{4 \pi} \right],
\qquad \rho_v < 1                                       \label{enFreeS} %
\eeq
one can estimate that the surface barrier $U$ is destroyed
$$
\frac{dF}{d\rho_v} \simeq \frac{1}{16 \pi} %
- \frac{\rho_v H_0}{4} \leq 0
\quad {\rm for} \quad \rho_v=\xi,                       \label{dGibbsS} %
$$
when the applied magnetic field $H_0$ becomes of the order of the
thermodynamic critical field in the superconductor
$H_{cm} = \frac{\ds \kappa}{\ds 2 \pi\, \sqrt{2}}$.
Thus, the same applied magnetic fields are required to % completely
destroy both the surface barrier for a vortex half--ring and
the Bean--Livingston barrier
for a rectilinear VL parallel to the surface.
%
%
%
%%%%%%%%%%%%%%%%%%%%%%%%%%%%%%%%%%%%%%%%%%%%%%%%%%%%%%%%%%%%%%%%%%%%%%
%                    4. DISCUSSION OF THE RESULTS                    %
%%%%%%%%%%%%%%%%%%%%%%%%%%%%%%%%%%%%%%%%%%%%%%%%%%%%%%%%%%%%%%%%%%%%%%
\vspace{10mm}
\begin{center}
{\small\bf 4. DISCUSSION OF THE RESULTS }
\end{center}

The solutions of the London equations were obtained here to describe
the structure of a curved Abrikosov vortex near the flat surface
of a type--II superconductor.
The VL of such a vortex is a semicircle of arbitrary radius.
The interaction of this vortex half--ring and an applied magnetic field
was considered by calculating the Gibbs free energy.
This problem extends the well--known ideas concerning the surface
Bean--Livingston barrier %
\cite{Bean64},
for the case where the Abrikosov vortex is curved.
Vortex loops like these represent an example of compact magnetic
structures in SC
\cite{Williams94}.
The specific feature of such magnetic structures is the self--consistent
localization of both the supercurrent and the magnetic field in all three
dimensions.
Due to the continuity of the magnetic flux, the end points of the vortex
loop touch the surface of SC, and the vortex line produces
a scattered field ${\bf H}_s$ outside the superconductor.
The creation of the vortex loops is depressed significantly by
the surface energy barrier.
The height of this barrier is reduced by increasing of the applied
magnetic field $_0$.
The surface barrier disappears completely when the field $_0$
becomes of the order of the thermodynamic critical field $H_{cm}$, and
the deterministic nucleation of small vortex loops with
size comparable to the coherence length $\xi$ is possible.
The applied magnetic field ${\bf H}_0$ produces the Meissner
supercurrent
\beq
{\bf j}_{\it m} = \frac{1}{4 \pi} \rot{\bf H}_{\it m},  \label{jMeissner} %
\eeq
distributed mainly in the layer of thickness $\lambda$ near
the surface.
At the surface the density of the Meissner current, corresponding
to the applied field $H_0 = H_{cm}$, is comparable to the
depairing current density
$j_c = \frac{\ds \kappa}{\ds 12 \sqrt{3}\, \pi^2}$
\cite{Abrikosov&Fund}.
Since the inhomogeneity of the Meissner current density is not
important at scales of the coherence length $\xi$,
the condition
$j_m \sim j_c$ determines the threshold of thermodynamic stability
of the uniform current carrying state in superconductors
\cite{Galaiko66,Abrikosov&Fund}
and agrees with the Landau criterion of roton excitation in a
superfluid flow
\cite{Kolomeisky91,Feynman&Stat}.

At a nonzero temperature $T$, before the surface barrier
is destroyed ($H_0 < H_{cm}$), the penetration of magnetic
flux into the SC occurs due to thermally activated nucleation,
subsequent growth, and merging of vortex half--rings.
A similar mechanism of thermally assisted overcoming
of energy barriers should be important at high temperatures
$T$ near $T_c$
\cite{Burlachkov94} %
especially in materials with strong thermal fluctuations
\cite{Fisher91,Nelson89}.

The creation of expanding vortex loops near a
superconductor surface is similar to the generation of
closed Abrikosov vortices by an external current
\cite{Fisher91,Kolomeisky91}.
Since the energy of the vortex half--ring $G_v$ (\ref{enTorus})
grows monotonically with the radius of the VL $\rho_v$,
the nucleation and expansion of vortex loops results in the
increase of the total free energy of superconductor.
The decrease of the Gibbs free energy $F$ (\ref{enGibbsEx})
is determined by the work  $\Delta W(\rho_v)$, performed by the
source of the applied magnetic field ${\bf H}_0$ during
the expansion of the vortex loop up to size $\rho_v$:
%\cite{Campbell&Crit,Abrikosov&Fund}:
%
\beq
\Delta W(\rho_v) = \frac{H_0 \rho_v}{2 \pi}\; %
\Bigl[ 1-\int\limits_0^{\pi/2} d\varphi \sin{\varphi} %
        \exp{(-\rho_v \sin\varphi)} \Bigr].             \label{sourcework} %
\eeq
The work $\Delta W(\rho_v)$ is related to the motion of the
vortex line under the action of the Lorentz force $f_L$
\cite{Schmidt74}
\beq
{\bf f}_{\it L} = \left[ {\bf j}_{\it m} %
\times {\bf e}_{\it v} \right]                          \label{fLorentz} %
\eeq
caused by the Meissner current ${\bf j}_m$.
Since the current ${\bf j}_m$ decreases outside a surface layer
of thickness $\sim \lambda$, the interaction of the applied
magnetic field and the vortex half--ring depends strongly
on the size of the vortex $\rho_v$.
If $\rho_v \ll 1$, the vortex line is located wholly in the
area where the current density ${\bf j}_m$ is distributed uniformly,
and $\Delta W \sim {\rho_v}^2$
\cite{Fisher91,Kolomeisky91,Chen93}.
In the other limiting case ($\rho_v \gg 1$),
the integrand in the expression (\ref{sourcework})
drops abruptly if $\sin{\varphi} \sim 1$,
and the region $\sin{\varphi} \ltsim 1/ \rho_v \ll 1$
makes an essential contribution
to the integral in (\ref{sourcework}).
Replacing
$\sin{\varphi} \simeq \varphi$
in (\ref{sourcework})
it is easy to obtain
$$
\Delta W(\rho_v) \simeq \frac{H_0 \rho_v}{2 \pi}
\quad {\rm for} \quad \rho_v \gg 1.
$$
In this case the basic part of
the vortex line is located outside the
surface layer of a thickness $\sim \lambda$,
where the Meissner
current ${\bf j}_m$ is absent,
and the work $\delta W$ is produced by the motion
of parts of the VL, that border on the surface.

The equilibrium shape of the Abrikosov vortex is determined mainly
by the structure of the applied magnetic field and represents
a rectilinear VL parallel to the surface.
The semicircle is the correct form of a vortex nucleus
near the surface of the superconductor only if the vortex radius
$\rho_v \ll 1$.
In this case the Meissner current ${\bf j}_m$ does not change essentially
on scales of the size of the vortex, so the value of the Lorentz force
$f_L$ (\ref{fLorentz}) is the same for different points of the vortex
line.
Thus an expansion of the vortex half--ring takes place without
distortions of the form of the vortex line.
When the size of the vortex $\rho_v$ becomes comparable
to the magnetic field penetration depth $\lambda$
($\rho_v \gtrsim 1$), the value of the Lorentz force
$f_L$ reduces strongly for the parts of the vortex line, placed
farther from the surface.
It results in distortions of the form of the vortex line
so that vortex loops elongate parallel to the
surface.
It is evident that the vortex line of such an elongated vortex
loop has a large part directed along the applied
magnetic field.

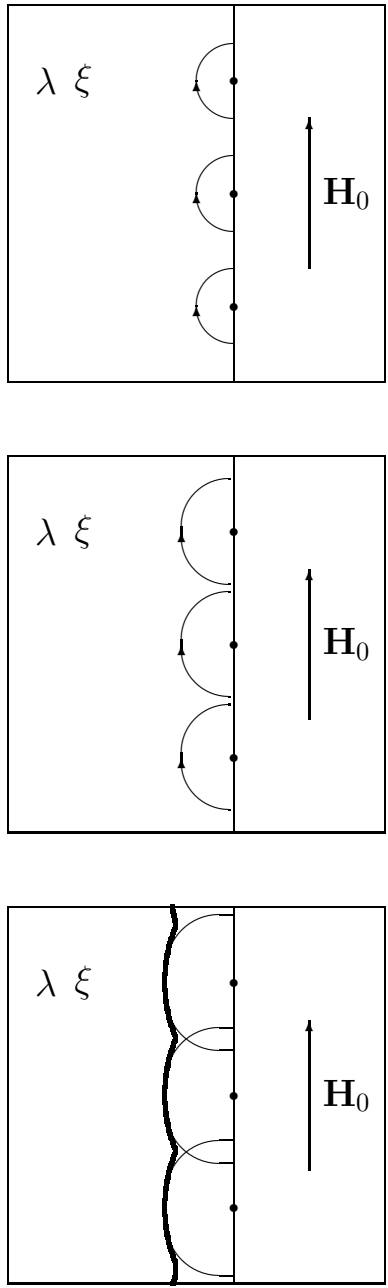
\begin{figure}[htb]
\unitlength=1.00mm
\special{em:linewidth 0.4pt}
\linethickness{0.4pt}
\begin{picture}(100.00,190.00)
\put(50.00,140.00){\framebox(50.00,50.00)[cc]{}}
\put(80.00,190.00){\line(0,-1){0.06}}
\put(79.94,164.94){\circle*{1.01}}
\put(79.94,179.94){\circle*{1.01}}
\put(79.94,149.94){\circle*{1.01}}
\put(80.00,180.00){\oval(10.00,10.00)[l]}
\put(80.00,165.00){\oval(10.00,10.00)[l]}
\put(80.00,150.00){\oval(10.00,10.00)[l]}
\put(90.00,155.00){\vector(0,1){20.00}}
\put(75.00,180.00){\vector(0,0){0.03}}
\put(75.00,165.00){\vector(0,0){0.03}}
\put(75.00,150.00){\vector(0,0){0.03}}
\put(95.00,165.00){\makebox(0,0)[cc]{${\bf H}_0$}}
\put(55.00,180.00){\makebox(0,0)[cc]{$\lambda$}}
\put(60.00,180.00){\makebox(0,0)[cc]{$\xi$}}
\put(80.00,140.00){\line(0,1){0.04}}
\put(80.00,140.00){\line(0,1){50.00}}
\put(80.00,190.00){\line(0,0){0.00}}
\put(50.00,80.00){\framebox(50.00,50.00)[cc]{}}
\put(80.00,130.00){\line(0,-1){0.06}}
\put(79.94,104.94){\circle*{1.01}}
\put(79.94,119.94){\circle*{1.01}}
\put(79.94,89.94){\circle*{1.01}}
\put(90.00,95.00){\vector(0,1){20.00}}
\put(95.00,105.00){\makebox(0,0)[cc]{${\bf H}_0$}}
\put(55.00,120.00){\makebox(0,0)[cc]{$\lambda$}}
\put(60.00,120.00){\makebox(0,0)[cc]{$\xi$}}
\put(80.00,80.00){\line(0,1){0.04}}
\put(80.00,80.00){\line(0,1){50.00}}
\put(80.00,130.00){\line(0,0){0.00}}
\put(79.50,120.00){\oval(13.00,14.00)[l]}
\put(79.50,105.00){\oval(13.00,14.00)[l]}
\put(79.50,90.00){\oval(13.00,14.00)[l]}
\put(73.00,120.00){\vector(0,0){0.01}}
\put(73.00,105.00){\vector(0,0){0.01}}
\put(73.00,90.00){\vector(0,0){0.01}}
\put(50.00,20.00){\framebox(50.00,50.00)[cc]{}}
\put(80.00,70.00){\line(0,-1){0.06}}
\put(79.94,44.94){\circle*{1.01}}
\put(79.94,59.94){\circle*{1.01}}
\put(79.94,29.94){\circle*{1.01}}
\put(90.00,35.00){\vector(0,1){20.00}}
\put(95.00,45.00){\makebox(0,0)[cc]{${\bf H}_0$}}
\put(55.00,60.00){\makebox(0,0)[cc]{$\lambda$}}
\put(60.00,60.00){\makebox(0,0)[cc]{$\xi$}}
\put(80.00,20.00){\line(0,1){0.04}}
\put(80.00,20.00){\line(0,1){50.00}}
\put(80.00,70.00){\line(0,0){0.00}}
\linethickness{0.25pt}
\put(80.00,60.00){\oval(18.00,18.00)[l]}
\put(80.00,45.00){\oval(18.00,18.00)[l]}
\put(80.00,30.00){\oval(18.00,18.00)[l]}
\linethickness{1.5pt}
\bezier{64}(71.99,67.57)(69.02,60.02)(71.99,52.60)
\bezier{64}(71.99,52.57)(69.02,45.02)(71.99,37.60)
\bezier{64}(71.99,37.57)(69.02,30.02)(71.99,22.60)
\bezier{12}(71.99,67.57)(71.74,68.93)(71.49,70.04)
\bezier{12}(71.99,22.42)(71.99,20.57)(71.61,19.95)
%\put(70.50,59.26){\vector(0,1){1.55}}
%\put(70.50,44.08){\vector(0,1){1.79}}
%\put(70.50,29.26){\vector(0,1){1.79}}
\end{picture}
\caption{The formation of the rectilinear Abrikosov vortex,
parallel to the surface, by intersection
and cross joining vortex loops. } %
\label{Crossing}
\end{figure}

The pattern of expanding single vortex half--rings holds as long as
the radius of the vortex $\rho_v$ is much smaller
than the typical intervortex distance $L$.
If the intervortex distance
becomes comparable with the sizes of the vortex,
the regular expansion of vortex half--ring is changed by the
interaction of adjacent vortices.
The attraction of the sections of the vortices of
opposite directions may lead to intersection and cross joining
of the adjacent vortices
\cite{Campbell&Crit}.
Figure~\ref{Crossing} shows the sequence of formation of a
rectilinear Abrikosov vortex parallel to the surface by
joining of expanding vortex loops.
It is essential, that if the applied magnetic field is not
enough to destroy the surface barrier ($H_0 < H_{cm}$),
and the thermally activated motion of vortices over the barrier
takes place, the creation of a rectilinear Abrikosov
vortex near the surface by intersection and joining of the
vortex loops can allow for the magnetic field entry in type--II
superconductor.
The surface barrier in this case is determined by the energy barrier
$U$ preventing generation of a single vortex half--ring
and may be lower than the ordinary BL barrier for
a rectilinear Abrikosov vortex
\cite{Bean64}.
The adjacent vortex loops, which penetrate after the first one,
require less energy than $U$
\cite{Burlachkov94}.
Therefore, the process of the formation of a rectilinear Abrikosov
vortex is of avalanche--type, and $U$ is the activation barrier
for the vortex entry
\cite{Mints93}.

%
%
%%%%%%%%%%%%%%%%%%%%%%%%%%%%%%%%%%%%%%%%%%%%%%%%%%%%%%%%%%%%%%%%%%%%%%
%                       CONCLUSION                                   %
%%%%%%%%%%%%%%%%%%%%%%%%%%%%%%%%%%%%%%%%%%%%%%%%%%%%%%%%%%%%%%%%%%%%%%
\vspace{10mm}
\begin{center}
{\small\bf 5. CONCLUSION}
\end{center}

In this paper the properties of the Abrikosov vortex in the form of
a semicircle of arbitrary radius $\rho_v$ near the flat
surface of type--II superconductor in the applied magnetic
field ${\bf H}_0$, parallel to the surface, have been
theoretically studied.
The magnetic field structure, the magnetic flux and the free energy
of a vortex half--ring have been calculated in the London approximation.
The creation of such vortices is delayed by the surface barrier,
similar to the Bean--Livingston barrier for a rectilinear Abrikosov
vortex
\cite{Bean64}.
The barrier is reduced by increasing the applied magnetic field
$H_0$ and vanishes in the thermodynamic critical field $H_{cm}$,
so that the deterministic nucleation of vortex loops with the size
of coherence length $\xi$ becomes possible.
At temperatures $T$ near $T_c$ and for applied fields
$H_0 < H_{cm}$ the magnetic flux penetrates into a type--II
superconductor by thermally activated nucleation of expanding
vortex half--rings near the surface.
The formation of the rectilinear Abrikosov vortices by
intersection and cross joining of vortex loops has been
considered.
\newpage
%%%%%%%%%%%%%%%%%%%%%%%%%%%%%%%%%%%%%%%%%%%%%%%%%%%%%%%%%%%%%%%%%%%%%%
%                       APPENDIX A                                   %
%%%%%%%%%%%%%%%%%%%%%%%%%%%%%%%%%%%%%%%%%%%%%%%%%%%%%%%%%%%%%%%%%%%%%%
%\vspace{10mm}
\begin{center}
{\small\bf APPENDIX A}
\end{center}

Let us express the spectral function $S_v(q,u)$ by means of
the auxiliary function $w_{qu}(z)$, satisfying the following
inhomogeneous differential equation with the zero initial
conditions:
$$
w_{qu}^{''} - \frac{1}{z} w_{qu}^{'} %
- (1+q^2) w_{qu} = - \sin{(uz)},                        \eqno({\rm A.1})
$$
$$
w_{qu}(0) = w_{qu}^{'}(0) = 0.
$$
Writing the Green function for equation (A.1), one can obtain the
particular solution:
\beqr
& &w_{qu}(z) = z K_1(z \sqrt{\ds 1+q^2}) %
\int\limits_0^z d \zeta\, \sin{(u \zeta)}\,
I_1(\zeta \sqrt{\ds 1+q^2}) \nonumber \\ %
& &\qquad\quad - z I_1(z \sqrt{\ds 1+q^2}) %
\int\limits_0^z d\zeta\, \sin{(u \zeta)}\,
K_1(\zeta \sqrt{\ds 1+q^2}). \nonumber
\eeqr
Using the known integral:
$$
\int\limits_0^{+\infty} d \zeta\,K_1(a \zeta)\,\sin{(u \zeta)} = %
\frac{\pi u}{2 a \sqrt{\ds a^2+u^2}},
$$
one can set the following relation between the functions $S_v(q,u)$
and $w_{qu}(z)$:
$$
S_v(q,u) = w_{qu}(\rho_v) + %
\frac{\pi u \rho_v I_1(\rho_v \sqrt{\ds 1+q^2})} %
     {2 \sqrt{\ds 1+q^2}\,\sqrt{\ds 1+q^2+u^2}}.        \eqno({\rm A.2}) %
$$
Thus, to simulate the value of the function $S_v(q,u)$ for
arbitrary $q$ and $u$ it is possible to solve the equation (A.1)
on the interval $z \in [0,\rho_v]$ and to substitute the
obtained value $w_{qu}(\rho_v)$ in expression (A.2).
This method of calculation of the spectral function $S_v(q,u)$
is especially efficient for large values $q,u \gg 1$, because it
uses the well--known algorithms of solution of Cauchy problem
instead of simulations of improper integrals on rapidly
oscillating functions.
\newpage
%%%%%%%%%%%%%%%%%%%%%%%%%%%%%%%%%%%%%%%%%%%%%%%%%%%%%%%%%%%%%%%%%%%%%%
%                             APPENDIX B                             %
%%%%%%%%%%%%%%%%%%%%%%%%%%%%%%%%%%%%%%%%%%%%%%%%%%%%%%%%%%%%%%%%%%%%%%
%\vspace{10mm}
\begin{center}
{\small\bf APPENDIX B}
\end{center}

The potential $U_P$ of the dipole $P=2 \rho_v Q$ placed in the
plane $z=0$ as shown in the insert of fig.~\ref{dipole}
is described by the well known expression
\cite{Landau&Fields}:
$$
U_P = P \frac{y}{R^3}, %
      \qquad R={(x^2+y^2+z^2)}^{1/2} \gg \rho_v.        \eqno({\rm B.1}) %
$$
Let us represent the potential $U_P$ in the form of the two
dimensional Fourier transform over spatial harmonics:
$$
U_P = \frac{1}{2 \pi} %
\int \limits_{-\infty}^{+\infty} dq \exp{(i q x)} %
\int \limits_{-\infty}^{+\infty} du \exp{(i u y)} %
S_P(q,u,z).                                             \eqno({\rm B.2}) %
$$
The spatial spectrum $S_P$ obeys the inverse Fourier transform
and should be written as follows:
$$
S_P = \frac{1}{2 \pi} %
\int \limits_{-\infty}^{+\infty} dx \exp{(-i q x)} %
\int \limits_{-\infty}^{+\infty} dy \exp{(-i u y)} %
U_P(x,y,z).                                             \eqno({\rm B.3}) %
$$
Substitution of $U_P$ (B.1) in the expansion (B.3) leads to the
following expression of the spectrum $S_P$:
$$
S_P = - \frac{2 i P}{\pi} u %
\int \limits_0^{+\infty} d\zeta\, \cos{(u \zeta)}\, %
K_0(q\, \sqrt{z^2+\zeta^2}),                            \eqno({\rm B.4}) %
$$
where $K_0(\zeta)$ is the McDonald function.
Using the known integral:
$$
\int \limits_0^{+\infty} d\zeta\, \cos{(u \zeta)}\, %
K_0(q\, \sqrt{z^2+\zeta^2}) %
= \frac{\ds \pi\, \exp{(-z \sqrt{q^2+u^2} )}} %
       {\ds 2\, \sqrt{q^2+u^2}} %
$$
one can obtain:
$$
S_P(q,u,z) = -i P \frac{\ds u\, \exp{(-z \sqrt{q^2+u^2})} } %
                       {\sqrt{q^2+u^2}}.                \eqno({\rm B.5}) %
$$
Thus, the potential $U_P$ of the dipole $P$ at long distances
$R \gg \rho_v$ is determined by the expression:
$$
U_P = \frac{2 P}{\pi}
\int \limits_0^{+\infty} dq \cos{(q x)}
\int \limits_0^{+\infty} du
\frac{\ds u \sin{(u y)}}{\ds \sqrt{q^2+u^2}}
\exp{(-z \sqrt{q^2+u^2})}.                              \eqno({\rm B.6}) %
$$
\newpage
%%%%%%%%%%%%%%%%%%%%%%%%%%%%%%%%%%%%%%%%%%%%%%%%%%%%%%%%%%%%%%%%%%%%%%
%                             APPENDIX C                             %
%%%%%%%%%%%%%%%%%%%%%%%%%%%%%%%%%%%%%%%%%%%%%%%%%%%%%%%%%%%%%%%%%%%%%%
%\vspace{10mm}
\begin{center}
{\small\bf APPENDIX C}
\end{center}

To compute the magnetic field value at the vortex line center,
we use the other representation of the magnetic field distribution
$H_v(\rho,x)$ in a toroidal Abrikosov vortex
\cite{Kozlov91}:
$$
H_v(\rho,x) =\frac{\rho_v}{2} %
\int\limits_0^{+\infty} dq \frac{ q\, J_1(q \rho)\, J_1(q \rho_v)\, %
\exp( - \vert x \vert\, \sqrt{1+q^2} ) }
{ \sqrt{1+q^2} } ,                                      \eqno({\rm C.1}) %
$$
where $J_1(\zeta)$ is the Bessel function of the first kind.
It is easy to verify, that expression (C.1) is the exact solution
of equation (\ref{eqnLondonR}).
The integral in (C.1) is divergent for $\rho=\rho_v$, $x=0$, because
of the integrand function falls down too slowly.
To estimate the magnetic field value $H_v(\rho_v,0)$ let us
extract the divergent part of integrand function
$$
S_h(q,\rho) = \frac{ q\, J_1(\rho q)\, J_1(\rho_v q) } %
{ \sqrt{1+q^2} } .
$$
Using the asymptotic expansion of the Bessel function
$J_1(\zeta)$ for $\zeta \gg 1$,
one can obtain the following behavior of the function $S_h(q,\rho)$:
$$
S_h(q,\rho) \simeq
        \frac{ \cos[ q (\rho-\rho_v) ]} %
             { \pi \rho_v \sqrt{1+q^2}} %
      - \frac{ \sin(2 q \rho_v) }{ \pi q \rho_v}
        \quad {\rm for} \quad %
        q \gg max(1,\; 1/\rho_v,\; 1/\rho).             \eqno({\rm C.2}) %
$$
The first term in (C.2) results in the logarithmic divergence of
expression (C.1) if $\rho \to \rho_v$, $x \to 0$, because of
$$
\int\limits_0^{+\infty} dq %
        \frac{\cos[q (\rho-\rho_v)]}{\sqrt{1+q^2}} %
        = K_0( \vert \rho - \rho_v \vert ),             \eqno({\rm C.3}) %
$$
where $K_0(\zeta)$ is the McDonald function.
The value of the last integral have to be truncated in the usual way
at the distance $\xi$ from the VL center by setting
$\vert \rho - \rho_v \vert = \xi$ in (C.3),
so that the following representation of
the magnetic field value $H_v(\rho_v,0)$ at the VL center holds:
$$
H_v(\rho_v,0) = \frac{K_0(\xi) -\pi/2}{2 \pi} %
+ \frac{\rho_v}{2} %
  \int\limits_0^{+\infty} dq %
  \left[ %
        \frac{ q\, J_1^2(q \rho_v) }{ \sqrt{1+q^2} } %
      - \frac{ 1 }{ \pi \rho_v \sqrt{1+q^2}} %
      + \frac{ \sin(2 q \rho_v) }{ \pi q \rho_v } %
  \right].                                              \eqno({\rm C.4})
$$
The new integrand function in (C.4) falls down quickly enough
to provide the convergence of the improper integral
in expression (C.4).
\newpage
\thispagestyle{empty}
%%%%%%%%%%%%%%%%%%%%%%%%%%%%%%%%%%%%%%%%%%%%%%%%%%%%%%%%%%%%%%%%%%%%%%
%                       REFERENCES                                   %
%%%%%%%%%%%%%%%%%%%%%%%%%%%%%%%%%%%%%%%%%%%%%%%%%%%%%%%%%%%%%%%%%%%%%%
%\vspace{10mm}
\begin{center}
{\small\bf REFERENCES}
\end{center}
%

%%%%%%%%%%%%%%%%%%%%%%%%%%%%%%%%%%%%%%%%%%%%%%%%%%%%%%%%%%%%%%%%%%%%%
%
\newpage
%                           Figure 1
\large
%
%\vspace{-10cm}
%\begin{figure}[htb]
%
%%%%%%%%%%%%%%%%%%%%%%%%%%%%%%%%%%%%%%%%%%%%%%%%%%%%%%%%%%%%%%%%%%%%%
\newpage
\large
%                           Figure 2 of revised manuscript
%
%\vspace{15mm}
\begin{figure}[htb]
\unitlength=0.55mm
\special{em:linewidth 0.4pt}
\linethickness{0.4pt}
\begin{picture}(80.00,90.00)(-135,-210)
\put(30.00,30.00){\vector(2,1){45.11}}
\put(35.00,50.00){\vector(3,-2){37.00}}
\put(50.00,20.00){\vector(0,1){55.11}}
\linethickness{1.0pt}
\linethickness{0.4pt}
\put(25.00,5.00){\line(2,1){55.00}}
\put(80.00,33.00){\line(0,1){57.00}}
\put(80.00,90.00){\line(-5,-2){54.89}}
\put(25.11,68.00){\line(0,-1){62.89}}
\put(53.00,74.00){\makebox(0,0)[cc]{$y$}}
\put(73.00,47.00){\makebox(0,0)[cc]{$x$}}
\put(76.00,26.00){\makebox(0,0)[cc]{$z$}}
\put(30.00,13.00){\makebox(0,0)[cc]{$S$}}
\put(53.00,57.00){\makebox(0,0)[lc]{$+Q$}}
\put(53.00,23.00){\makebox(0,0)[lc]{$-Q$}}
\put(50.00,55.00){\line(-5,3){10.00}}
\put(50.00,25.00){\line(-5,3){10.00}}
\put(44.00,29.00){\vector(0,1){29.00}}
\put(44.00,58.00){\vector(0,-1){29.00}}
\put(37.00,42.00){\makebox(0,0)[cc]{$2\rho_v$}}
\put(50.00,55.00){\circle*{4.00}}
\put(50.00,25.00){\circle*{4.00}}
\end{picture}
\end{figure}
%
%%%%%%%%%%%%%%%%%%%%%%%%%%%%%%%%%%%%%%%%%%%%%%%%%%%%%%%%%%%%%%%%%%%%%
\newpage
\large
%                           Figure 3
%
%
\vspace{10mm}
%
%%%%%%%%%%%%%%%%%%%%%%%%%%%%%%%%%%%%%%%%%%%%%%%%%%%%%%%%%%%%%%%%%%%%%
\newpage
\large
%                           Figure 4
%
%\vspace{15mm}
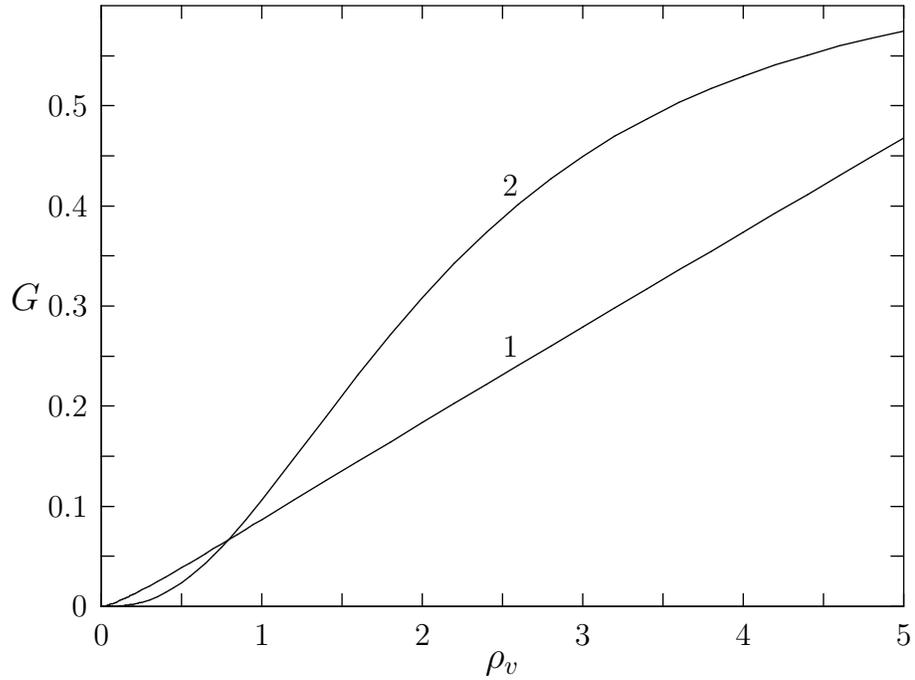
\begin{figure}[htb]
%\input{gv100gs.pic}
% GNUPLOT: LaTeX picture with emtex specials
\setlength{\unitlength}{0.240900pt}
\ifx\plotpoint\undefined\newsavebox{\plotpoint}\fi
\sbox{\plotpoint}{\rule[-0.500pt]{1.000pt}{1.000pt}}%
\special{em:linewidth 0.5pt}%
\begin{picture}(1500,1080)(0,0)
\font\gnuplot=cmr12 at 12pt
\gnuplot
\put(176,113){\special{em:moveto}}
\put(1436,113){\special{em:lineto}}
\put(176,113){\special{em:moveto}}
\put(176,1057){\special{em:lineto}}
\put(176,113){\special{em:moveto}}
\put(196,113){\special{em:lineto}}
\put(1436,113){\special{em:moveto}}
\put(1416,113){\special{em:lineto}}
\put(154,113){\makebox(0,0)[r]{0}}
\put(176,192){\special{em:moveto}}
\put(196,192){\special{em:lineto}}
\put(1436,192){\special{em:moveto}}
\put(1416,192){\special{em:lineto}}
\put(154,192){\makebox(0,0)[r]{ }}
\put(176,270){\special{em:moveto}}
\put(196,270){\special{em:lineto}}
\put(1436,270){\special{em:moveto}}
\put(1416,270){\special{em:lineto}}
\put(154,270){\makebox(0,0)[r]{0.1}}
\put(176,349){\special{em:moveto}}
\put(196,349){\special{em:lineto}}
\put(1436,349){\special{em:moveto}}
\put(1416,349){\special{em:lineto}}
\put(154,349){\makebox(0,0)[r]{ }}
\put(176,428){\special{em:moveto}}
\put(196,428){\special{em:lineto}}
\put(1436,428){\special{em:moveto}}
\put(1416,428){\special{em:lineto}}
\put(154,428){\makebox(0,0)[r]{0.2}}
\put(176,506){\special{em:moveto}}
\put(196,506){\special{em:lineto}}
\put(1436,506){\special{em:moveto}}
\put(1416,506){\special{em:lineto}}
\put(154,506){\makebox(0,0)[r]{ }}
\put(176,585){\special{em:moveto}}
\put(196,585){\special{em:lineto}}
\put(1436,585){\special{em:moveto}}
\put(1416,585){\special{em:lineto}}
\put(154,585){\makebox(0,0)[r]{0.3}}
\put(176,664){\special{em:moveto}}
\put(196,664){\special{em:lineto}}
\put(1436,664){\special{em:moveto}}
\put(1416,664){\special{em:lineto}}
\put(154,664){\makebox(0,0)[r]{ }}
\put(176,742){\special{em:moveto}}
\put(196,742){\special{em:lineto}}
\put(1436,742){\special{em:moveto}}
\put(1416,742){\special{em:lineto}}
\put(154,742){\makebox(0,0)[r]{0.4}}
\put(176,821){\special{em:moveto}}
\put(196,821){\special{em:lineto}}
\put(1436,821){\special{em:moveto}}
\put(1416,821){\special{em:lineto}}
\put(154,821){\makebox(0,0)[r]{ }}
\put(176,900){\special{em:moveto}}
\put(196,900){\special{em:lineto}}
\put(1436,900){\special{em:moveto}}
\put(1416,900){\special{em:lineto}}
\put(154,900){\makebox(0,0)[r]{0.5}}
\put(176,978){\special{em:moveto}}
\put(196,978){\special{em:lineto}}
\put(1436,978){\special{em:moveto}}
\put(1416,978){\special{em:lineto}}
\put(154,978){\makebox(0,0)[r]{ }}
\put(176,113){\special{em:moveto}}
\put(176,133){\special{em:lineto}}
\put(176,1057){\special{em:moveto}}
\put(176,1037){\special{em:lineto}}
\put(176,68){\makebox(0,0){0}}
\put(302,113){\special{em:moveto}}
\put(302,133){\special{em:lineto}}
\put(302,1057){\special{em:moveto}}
\put(302,1037){\special{em:lineto}}
\put(302,68){\makebox(0,0){ }}
\put(428,113){\special{em:moveto}}
\put(428,133){\special{em:lineto}}
\put(428,1057){\special{em:moveto}}
\put(428,1037){\special{em:lineto}}
\put(428,68){\makebox(0,0){1}}
\put(554,113){\special{em:moveto}}
\put(554,133){\special{em:lineto}}
\put(554,1057){\special{em:moveto}}
\put(554,1037){\special{em:lineto}}
\put(554,68){\makebox(0,0){ }}
\put(680,113){\special{em:moveto}}
\put(680,133){\special{em:lineto}}
\put(680,1057){\special{em:moveto}}
\put(680,1037){\special{em:lineto}}
\put(680,68){\makebox(0,0){2}}
\put(806,113){\special{em:moveto}}
\put(806,133){\special{em:lineto}}
\put(806,1057){\special{em:moveto}}
\put(806,1037){\special{em:lineto}}
\put(806,68){\makebox(0,0){ }}
\put(932,113){\special{em:moveto}}
\put(932,133){\special{em:lineto}}
\put(932,1057){\special{em:moveto}}
\put(932,1037){\special{em:lineto}}
\put(932,68){\makebox(0,0){3}}
\put(1058,113){\special{em:moveto}}
\put(1058,133){\special{em:lineto}}
\put(1058,1057){\special{em:moveto}}
\put(1058,1037){\special{em:lineto}}
\put(1058,68){\makebox(0,0){ }}
\put(1184,113){\special{em:moveto}}
\put(1184,133){\special{em:lineto}}
\put(1184,1057){\special{em:moveto}}
\put(1184,1037){\special{em:lineto}}
\put(1184,68){\makebox(0,0){4}}
\put(1310,113){\special{em:moveto}}
\put(1310,133){\special{em:lineto}}
\put(1310,1057){\special{em:moveto}}
\put(1310,1037){\special{em:lineto}}
\put(1310,68){\makebox(0,0){ }}
\put(1436,113){\special{em:moveto}}
\put(1436,133){\special{em:lineto}}
\put(1436,1057){\special{em:moveto}}
\put(1436,1037){\special{em:lineto}}
\put(1436,68){\makebox(0,0){5}}
\put(176,113){\special{em:moveto}}
\put(1436,113){\special{em:lineto}}
\put(1436,1057){\special{em:lineto}}
\put(176,1057){\special{em:lineto}}
\put(176,113){\special{em:lineto}}
\put(806,23){\makebox(0,0){$\rho_v$}}
\put(80,600){\makebox(0,0)[r]{$G$}}
\put(806,522){\makebox(0,0)[l]{1}}
\put(806,774){\makebox(0,0)[l]{2}}
\put(176,113){\special{em:moveto}}
\put(179,113){\special{em:lineto}}
\put(181,113){\special{em:lineto}}
\put(184,114){\special{em:lineto}}
\put(186,115){\special{em:lineto}}
\put(189,116){\special{em:lineto}}
\put(191,117){\special{em:lineto}}
\put(194,117){\special{em:lineto}}
\put(196,118){\special{em:lineto}}
\put(199,119){\special{em:lineto}}
\put(201,120){\special{em:lineto}}
\put(204,122){\special{em:lineto}}
\put(206,123){\special{em:lineto}}
\put(209,124){\special{em:lineto}}
\put(211,125){\special{em:lineto}}
\put(214,126){\special{em:lineto}}
\put(216,127){\special{em:lineto}}
\put(219,128){\special{em:lineto}}
\put(221,130){\special{em:lineto}}
\put(224,131){\special{em:lineto}}
\put(226,132){\special{em:lineto}}
\put(229,133){\special{em:lineto}}
\put(234,136){\special{em:lineto}}
\put(236,137){\special{em:lineto}}
\put(239,139){\special{em:lineto}}
\put(252,145){\special{em:lineto}}
\put(264,152){\special{em:lineto}}
\put(277,159){\special{em:lineto}}
\put(289,166){\special{em:lineto}}
\put(302,174){\special{em:lineto}}
\put(315,181){\special{em:lineto}}
\put(327,188){\special{em:lineto}}
\put(340,196){\special{em:lineto}}
\put(352,204){\special{em:lineto}}
\put(365,211){\special{em:lineto}}
\put(378,219){\special{em:lineto}}
\put(390,226){\special{em:lineto}}
\put(403,234){\special{em:lineto}}
\put(415,242){\special{em:lineto}}
\put(428,249){\special{em:lineto}}
\put(478,280){\special{em:lineto}}
\put(529,311){\special{em:lineto}}
\put(579,341){\special{em:lineto}}
\put(630,371){\special{em:lineto}}
\put(680,402){\special{em:lineto}}
\put(730,432){\special{em:lineto}}
\put(781,462){\special{em:lineto}}
\put(831,492){\special{em:lineto}}
\put(882,522){\special{em:lineto}}
\put(932,552){\special{em:lineto}}
\put(982,582){\special{em:lineto}}
\put(1033,612){\special{em:lineto}}
\put(1083,642){\special{em:lineto}}
\put(1134,671){\special{em:lineto}}
\put(1184,701){\special{em:lineto}}
\put(1234,731){\special{em:lineto}}
\put(1285,760){\special{em:lineto}}
\put(1335,790){\special{em:lineto}}
\put(1386,820){\special{em:lineto}}
\put(1436,849){\special{em:lineto}}
\put(176,113){\special{em:moveto}}
\put(179,113){\special{em:lineto}}
\put(181,113){\special{em:lineto}}
\put(184,113){\special{em:lineto}}
\put(186,113){\special{em:lineto}}
\put(189,113){\special{em:lineto}}
\put(191,113){\special{em:lineto}}
\put(194,113){\special{em:lineto}}
\put(196,113){\special{em:lineto}}
\put(199,113){\special{em:lineto}}
\put(201,113){\special{em:lineto}}
\put(204,114){\special{em:lineto}}
\put(206,114){\special{em:lineto}}
\put(209,114){\special{em:lineto}}
\put(211,114){\special{em:lineto}}
\put(214,115){\special{em:lineto}}
\put(216,115){\special{em:lineto}}
\put(219,115){\special{em:lineto}}
\put(221,116){\special{em:lineto}}
\put(224,116){\special{em:lineto}}
\put(226,116){\special{em:lineto}}
\put(229,117){\special{em:lineto}}
\put(234,118){\special{em:lineto}}
\put(236,119){\special{em:lineto}}
\put(239,119){\special{em:lineto}}
\put(252,123){\special{em:lineto}}
\put(264,128){\special{em:lineto}}
\put(277,135){\special{em:lineto}}
\put(289,142){\special{em:lineto}}
\put(302,150){\special{em:lineto}}
\put(315,160){\special{em:lineto}}
\put(327,170){\special{em:lineto}}
\put(340,181){\special{em:lineto}}
\put(352,193){\special{em:lineto}}
\put(365,206){\special{em:lineto}}
\put(378,220){\special{em:lineto}}
\put(390,234){\special{em:lineto}}
\put(403,249){\special{em:lineto}}
\put(415,264){\special{em:lineto}}
\put(428,280){\special{em:lineto}}
\put(478,345){\special{em:lineto}}
\put(529,411){\special{em:lineto}}
\put(579,477){\special{em:lineto}}
\put(630,540){\special{em:lineto}}
\put(680,598){\special{em:lineto}}
\put(730,652){\special{em:lineto}}
\put(781,701){\special{em:lineto}}
\put(831,745){\special{em:lineto}}
\put(882,785){\special{em:lineto}}
\put(932,820){\special{em:lineto}}
\put(982,852){\special{em:lineto}}
\put(1033,879){\special{em:lineto}}
\put(1083,905){\special{em:lineto}}
\put(1134,927){\special{em:lineto}}
\put(1184,946){\special{em:lineto}}
\put(1234,964){\special{em:lineto}}
\put(1285,979){\special{em:lineto}}
\put(1335,994){\special{em:lineto}}
\put(1386,1006){\special{em:lineto}}
\put(1436,1017){\special{em:lineto}}
\end{picture}
\caption{Free energy of half of toroidal Abrikosov vortex $G_v$ (1) %
and the contribution $G_s$ of the surface of superconductor $S$ to %
the free energy of the vortex half--ring (2) vs. the radius of the vortex %
line $\rho_v$ ($\kappa=100$). %
To maintain the scale of the vertical axis fixed, the value %
$100 \cdot G_s$  is used .}
\label{Freen}
\end{figure}
%
%%%%%%%%%%%%%%%%%%%%%%%%%%%%%%%%%%%%%%%%%%%%%%%%%%%%%%%%%%%%%%%%%%%%%
\newpage
\large
%                           Figure 5
%
%
%%%%%%%%%%%%%%%%%%%%%%%%%%%%%%%%%%%%%%%%%%%%%%%%%%%%%%%%%%%%%%%%%%%%%
\newpage
\large
%                          Figure 6
%
%\vspace{15mm}
%                           Figure 6(insert)
%\vspace{10mm}
\begin{figure}[htb]
%\input{gibbsin.pic}
% GNUPLOT: LaTeX picture with emtex specials
\setlength{\unitlength}{0.140900pt}
\ifx\plotpoint\undefined\newsavebox{\plotpoint}\fi
\sbox{\plotpoint}{\rule[-0.500pt]{1.000pt}{1.000pt}}%
\special{em:linewidth 0.5pt}%
\begin{picture}(1125,900)(-1300,-2750)
\font\gnuplot=cmr12 at 12pt
\gnuplot
\put(176,530){\special{em:moveto}}
\put(1061,530){\special{em:lineto}}
\put(176,113){\special{em:moveto}}
\put(176,877){\special{em:lineto}}
\put(176,724){\special{em:moveto}}
\put(196,724){\special{em:lineto}}
\put(1061,724){\special{em:moveto}}
\put(1041,724){\special{em:lineto}}
\put(154,724){\makebox(0,0)[r]{ }}
\put(176,113){\special{em:moveto}}
\put(1061,113){\special{em:lineto}}
\put(1061,877){\special{em:lineto}}
\put(176,877){\special{em:lineto}}
\put(176,113){\special{em:lineto}}
\put(618,23){\makebox(0,0){$\rho_v$}}
\put(-13,530){\makebox(0,0)[l]{$F$}}
\put(50,724){\makebox(0,0)[l]{$U$}}
\put(505,461){\makebox(0,0)[l]{$\rho_c$}}
\put(492,724){\vector(-1,0){291}}
\put(505,717){\vector(0,-1){187}}
\put(176,530){\special{em:moveto}}
\put(201,538){\special{em:lineto}}
\put(227,560){\special{em:lineto}}
\put(252,585){\special{em:lineto}}
\put(277,609){\special{em:lineto}}
\put(302,632){\special{em:lineto}}
\put(328,652){\special{em:lineto}}
\put(353,671){\special{em:lineto}}
\put(378,687){\special{em:lineto}}
\put(404,701){\special{em:lineto}}
\put(429,712){\special{em:lineto}}
\put(454,720){\special{em:lineto}}
\put(479,725){\special{em:lineto}}
\put(505,727){\special{em:lineto}}
\put(530,726){\special{em:lineto}}
\put(555,723){\special{em:lineto}}
\put(581,717){\special{em:lineto}}
\put(606,707){\special{em:lineto}}
\put(631,696){\special{em:lineto}}
\put(656,681){\special{em:lineto}}
\put(682,664){\special{em:lineto}}
\put(707,643){\special{em:lineto}}
\put(732,621){\special{em:lineto}}
\put(758,596){\special{em:lineto}}
\put(783,569){\special{em:lineto}}
\put(808,539){\special{em:lineto}}
\put(833,507){\special{em:lineto}}
\put(859,471){\special{em:lineto}}
\put(884,434){\special{em:lineto}}
\put(909,395){\special{em:lineto}}
\put(935,353){\special{em:lineto}}
\put(960,309){\special{em:lineto}}
\put(985,263){\special{em:lineto}}
\put(1010,215){\special{em:lineto}}
\put(1036,165){\special{em:lineto}}
\put(1061,113){\special{em:lineto}}
\end{picture}
\end{figure}
%
%%%%%%%%%%%%%%%%%%%%%%%%%%%%%%%%%%%%%%%%%%%%%%%%%%%%%%%%%%%%%%%%%%%%%
\newpage
\large
%                           Figure 7
%
\vspace{15mm}
%
%%%%%%%%%%%%%%%%%%%%%%%%%%%%%%%%%%%%%%%%%%%%%%%%%%%%%%%%%%%%%%%%%%%%%
%
%
\end{document}